\documentclass[pre,amsmath,twocolumn]{revtex4}
\usepackage{bm}
\usepackage{amssymb}
\usepackage{graphicx}
\usepackage{subfigure}

\begin{document}

\title{Hydrodynamics in kinetically-constrained lattice-gas models}

\author{Eial Teomy$^{1,2}$}
\email{eialteom@gmail.com}
%\author{Paul Krapivsky$^{3}$}
%\email{paulk@bu.edu}
%\author{Baruch Meerson$^{4}$}
%\email{meerson@mail.huji.ac.il}
\author{Yair Shokef$^{1}$}
\email{shokef@tau.ac.il}

\affiliation{$^{1}$School of Mechanical Engineering and Sackler Center for Computational Molecular and Materials Science, Tel Aviv University, Tel Aviv 69978, Israel}
\affiliation{$^{2}$Department of Physics, Bar Ilan University, Ramat Gan 52900, Israel} 
%\affiliation{$^{3}$Department of Physics, Boston University, Boston, Massachusetts 02215, USA} 
%\affiliation{$^{4}$Racah Institute of Physics, Hebrew University of Jerusalem, Jerusalem 91904, Israel}

\begin{abstract}

Kinetically-constrained models are lattice-gas models that are used for describing glassy systems. By construction, their equilibrium state is trivial and there are no equal-time correlations between the occupancy of different sites. We drive such models out of equilibrium by connecting them to two reservoirs of different densities, and measure the response of the system to this perturbation. We find that under the proper coarse-graining, the behavior of these models may be expressed by a nonlinear diffusion equation, with a model- and density-dependent diffusion coefficient. We find a simple approximation for the diffusion coefficient, and show that the relatively mild discrepancy between the approximation and our numerical results arises due to non-negligible correlations that appear as the system is driven out of equilibrium, even when the density gradient is infinitesimally small. Similar correlations appear when such kinetically-constrained models are driven out of equilibrium by applying a uniform external force. We suggest that these correlations are the reason for the same discrepancy between the approximate diffusion coefficient and the numerical results for a broader group of models, non-gradient lattice-gas models, for which kinetically-constrained models are arguably the simplest example thereof.

\end{abstract}

\date{\today}
 
\maketitle

\section{Introduction}

Kinetically constrained models (KCMs) are a family of lattice-gas models designed to investigate glass-forming liquids \cite{017Ritort2003,018Garrahan2010}. By construction, the equilibrium state of these models is trivial. However their dynamics are cooperatively slow and they exhibit many hallmarks of glassy systems, such as dynamical heterogeneities \cite{100Berthier2003a,Chandler2006,055Leonard2010,106Pastore2013,Pastore2015a,Garrahan2002a,101Berthier2003,104Marinari2005,Whitelam2005,105Berthier2007,087Teomy2015,Pastore2016}, non-exponential relaxation \cite{114Butler1991,067Kob1993,094Einax2001,099Schulz2002,Garrahan2002a,101Berthier2003,Whitelam2005,095Kuhlmann2005,102Jung2005,104Marinari2005,105Berthier2007,087Teomy2015,Pastore2016}, and ageing \cite{Kurchan1997,097Leonard2007,098Mayer2007}, and in certain situations may exhibit an ergodicity-breaking jamming transition, beyond which a finite fraction of the particles are permanently frozen \cite{083Toninelli2007,084Biroli2008,115Teomy2012,082Teomy2014,085Ghosh2014}.

Most of the research on KCMs has focused on relaxation processes within the equilibrium state, however there are also several works on KCMs out of equilibrium. Such works investigated out of equilibrium systems relaxing to equilibrium \cite{Corberi2009}, and systems driven out of equilibrium by applying an external field \cite{Fielding2002,Sellitto2008,041Shokef2010,130Turci2012a} and by connecting them to external reservoirs \cite{126Sellitto2002,127Goncalves2009}. A different work considered spin diffusion in a heterogeneous KCM with spin-spin interactions and a kinetic-constraint that depends on the entire system \cite{Karabanov2015}.

In KCMs there are only hard-core interactions between particles, meaning that each site on the lattice can be occupied by at most one particle. However, the hopping rate of a particle to an adjacent vacant site depends on the configuration of the neighboring sites. For example, in the Kob-Andersen (KA) \cite{067Kob1993} model on a $d$-dimensional hypercubic lattice, a particle can hop to an adjacent vacant site if at least $m$ of its $2d$ nearest neighbors are vacant both before and after the move, and thus the hopping rate in this model is either $0$ if the move is not allowed, or some constant if the move is allowed. The simple symmetric exclusion principle (SSEP) model \cite{068Spitzer1970} is recovered in the case $m=1$. Because there are only hard-core interactions, in equilibrium there are no correlations between the occupancy of sites at the same time, and each site is independently occupied with probability $\rho$ and vacant with probability $1-\rho$. Of course, there are correlations between the occupancy of different sites at different times.

\begin{figure}
\begin{centering}
\includegraphics[width=0.8\columnwidth]{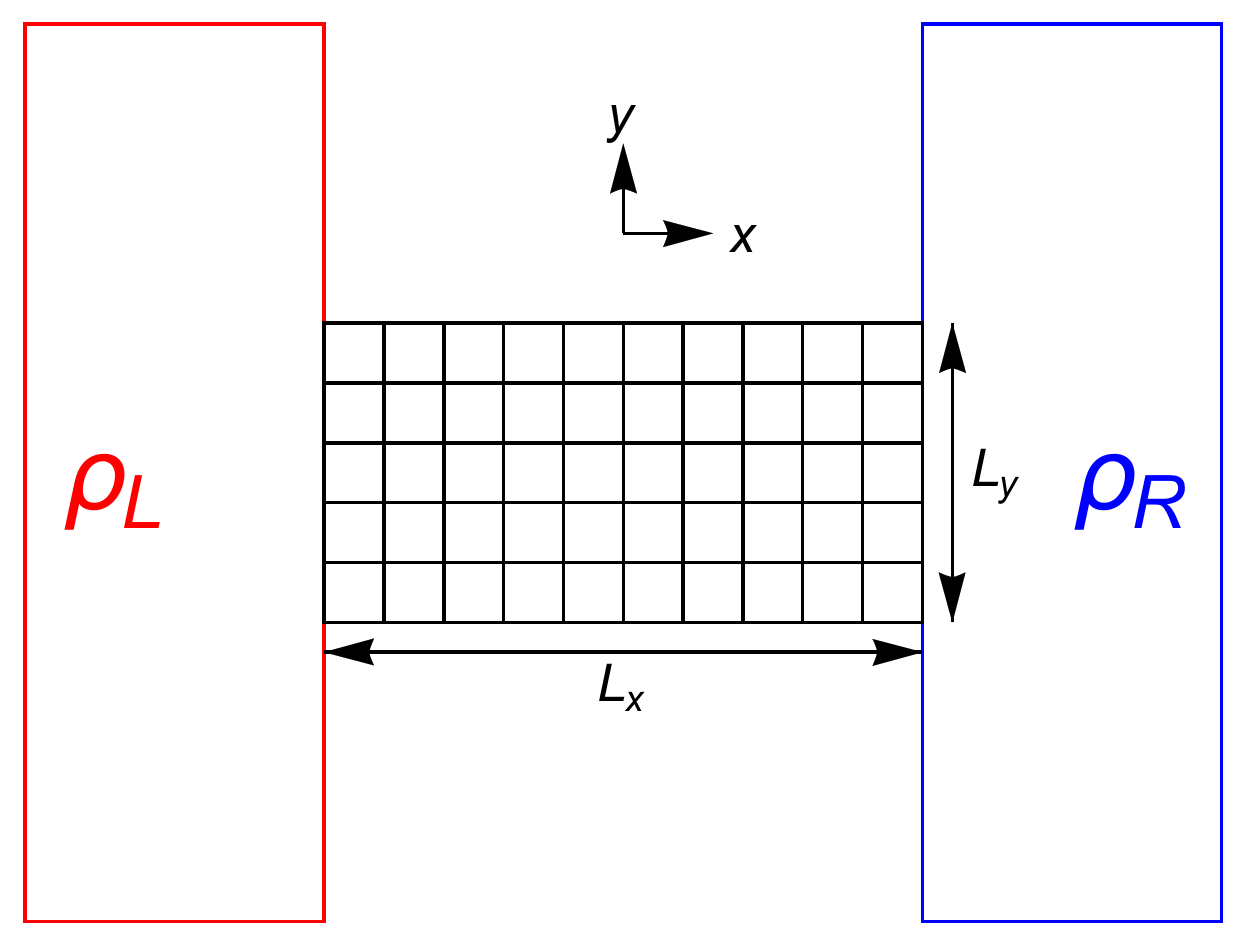}
\par\end{centering}
\caption{A sketch of the setup we consider here. The lattice in the middle is a system of $L_{x}=10$ by $L_{y}=5$ sites, which is connected to two particle reservoirs with two different densities $\rho_{L}$ and $\rho_{R}$. In the $y$ direction we employ periodic boundary conditions.}
\label{diffusionsketch}
\end{figure}

Here we investigate what happens when the system is driven out of equilibrium by connecting it to reservoirs with different densities, see Fig. \ref{diffusionsketch}. In such a setup, the local density gradient $\nabla\rho$ creates a current of particles in the system, $J$, which for weak gradients should scale linearly with $\nabla \rho$. Thus we can infer the diffusion coefficient, $D$, by Fick's law
\begin{align}
J=-D\nabla\rho .
\end{align}

Such setups have been investigated before in the $m=3$ KA model in three dimensions~\cite{126Sellitto2002}, and in a one-dimensional KCM in which a particle can move to an adjacent vacant site if it has at least two neighboring vacant sites either before or after the move~\cite{127Goncalves2009}, which may be thought of as the $m=1\frac{1}{2}$ KA model. In~\cite{126Sellitto2002} an approximation for the diffusion coefficient was found using the assumption that there is a critical density at $\rho_{c}=0.881$. However, this assumption was later proven wrong~\cite{071Toninelli2004a} and shown to be a finite-size effect. In~\cite{127Goncalves2009} a gradient lattice model with noncooperative dynamics was considered (see definition of gradient model below). These two properties made the derivation of the diffusion coefficient tractable.

\begin{figure}
\includegraphics[width=0.6\columnwidth]{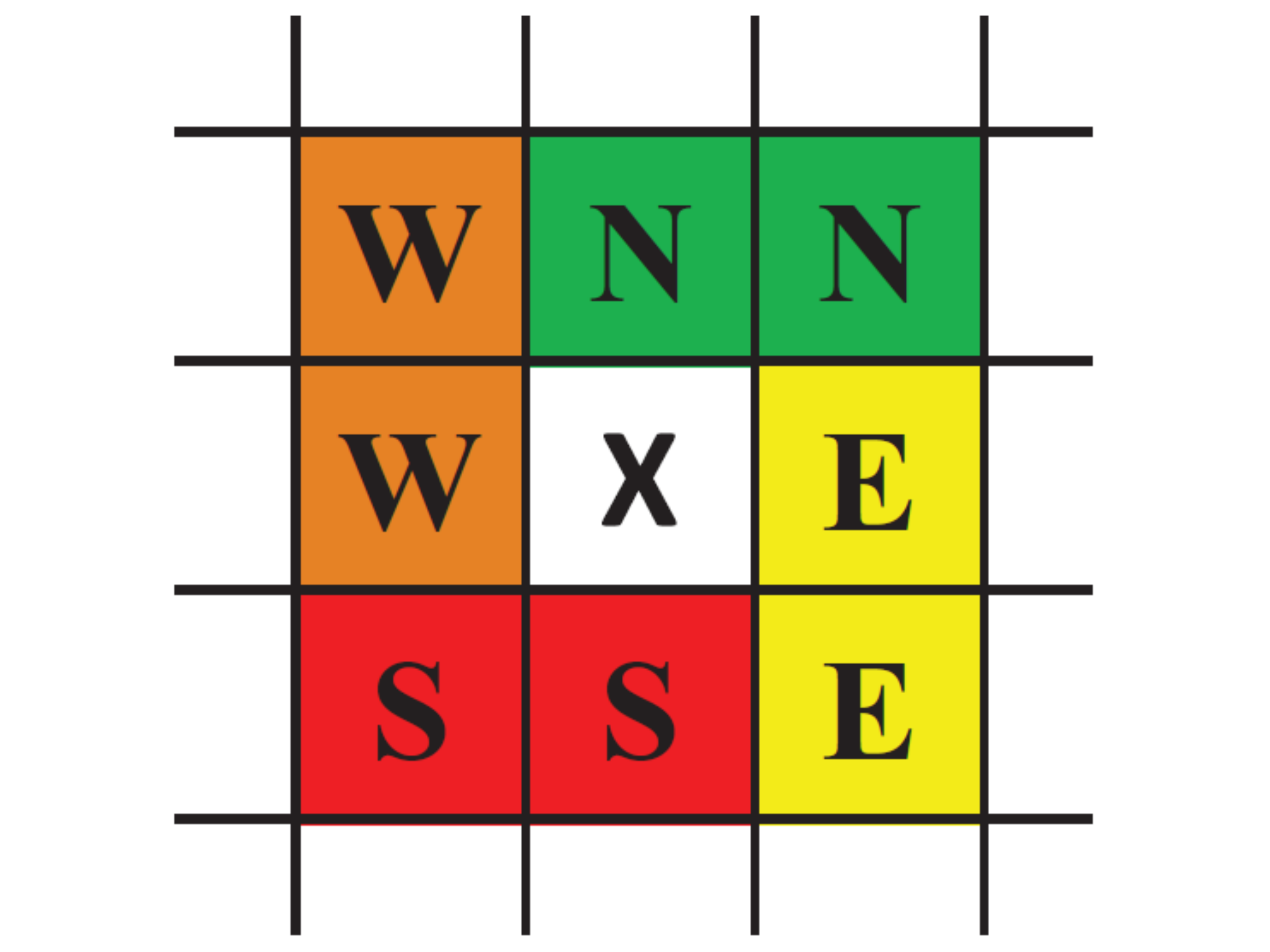}
\caption{The kinetic rules for the spiral model \cite{083Toninelli2007,084Biroli2008}. The eight neighbors of each site on the square lattice are divided into four groups: North, South, East and West. A site $(x)$ is \textit{unblocked} if either its North or South group is completely vacant and either its East or West group are completely vacant. A particle can move to an adjacent vacant site if it is in an unblocked site both before and after the move \cite{041Shokef2010}.}
\label{spiral_rules}
\end{figure}

In this paper we first consider KCMs in general and then concentrate on two specific models as examples: the KA model in two-dimensions and the spiral model \cite{083Toninelli2007,084Biroli2008}, the kinetic rules of which are shown in Fig. \ref{spiral_rules}. We will show that under the proper coarse-graining, the average particle density satisfies a nonlinear diffusion equation
\begin{align}
\frac{\partial\rho}{\partial t}=\frac{\partial}{\partial x}\left[D\left(\rho\right)\frac{\partial\rho}{\partial x}\right] ,\label{diffeq}
\end{align}
with a model-dependent, density dependent diffusion coefficient $D\left(\rho\right)$. We derive a simple approximation for the diffusion coefficient in general KCMs and show that the origin for the discrepancy between the simple approximation and the simulation results is non-negligible correlations that appear in the system even for an infinitesimally small perturbation out of equilibrium.

We note here that lattice gas models can be divided into two groups: gradient models and non-gradient models, where in gradient models the current may be written as a discrete gradient of some function of the density, and in non-gradient models it may not \cite{125Arita2014}. In gradient models, the expression we derive here for the diffusion coefficient is an exact result. However, generally KCMs, such as the specific models we consider here, are non-gradient models and we therefore do not expect the derivation to perfectly agree with the numerical results. The correlations we find here are related to the correlations that appear when the system is driven out of equilibrium by applying an external field \cite{Sellitto2008,130Turci2012a}. Finally, we further suggest a method to analytically derive the diffusion coefficient exactly for general KCMs, which is very cumbersome and beyond the scope of this paper.

The remaining of the paper is organized as follows: In Section \ref{sec_self} we define in detail the system's setup and emphasize the difference between the bulk- and self-diffusion coefficients. In Section \ref{sec_nc} we present our approximate derivation for the diffusion coefficient. In Section \ref{secdenprof} we compare the derived expression to simulation results and demonstrate that the discrepancy is not a finite-size effect but a genuine difference. In Section \ref{sec_corr} we discuss the origin of this discrepancy: non-negligible correlations that appear only when the system is driven out of equilibrium. We further show why in gradient models, even if correlations develop, the derived expression of the diffusion coefficient is exact, and not an approximation. In Section \ref{sec_fluct} we consider the fluctuations in the current and relate them to the diffusion coefficient via the fluctuation-dissipation theorem. In Section \ref{sec_num} we numerically find the diffusion coefficient, and find a better, but still relatively simple, approximation for it by combining the analytical approximation with a numerical observation. Section \ref{sec_summary} summarizes the paper.

\section{Setup}
\label{sec_self}

In what follows we consider a two dimensional square lattice of size $L_{x}\times L_{y}$ and will measure distances in units of the lattice constant. In the $y$-direction we consider periodic boundary conditions. In the $x$-direction the system is connected to reservoirs with densities $\rho_{L}$ and $\rho_{R}$, such that at all times the sites with $x$-coordinates $L_{x}+1$ or $L_{x}+2$ ($0$ or $-1$) are occupied with probability $\rho_{R}$ ($\rho_{L}$) independently of the occupancy of all other sites. As stated before, this implies delta-correlations between the occupancy of sites in the reservoirs, both in space and in time, which is very different from the dynamics of the sites inside the system which at least for the same site are obviously highly correlated in time. Moreover, in the non-equilibrium situation that we will consider, correlations develop between the occupancy of neighboring sites within the system. We further note that the kinetic constraints hold also for particles exiting the system and entering it from the reservoirs. 

\begin{figure}
\begin{centering}
\includegraphics[width=0.8\columnwidth]{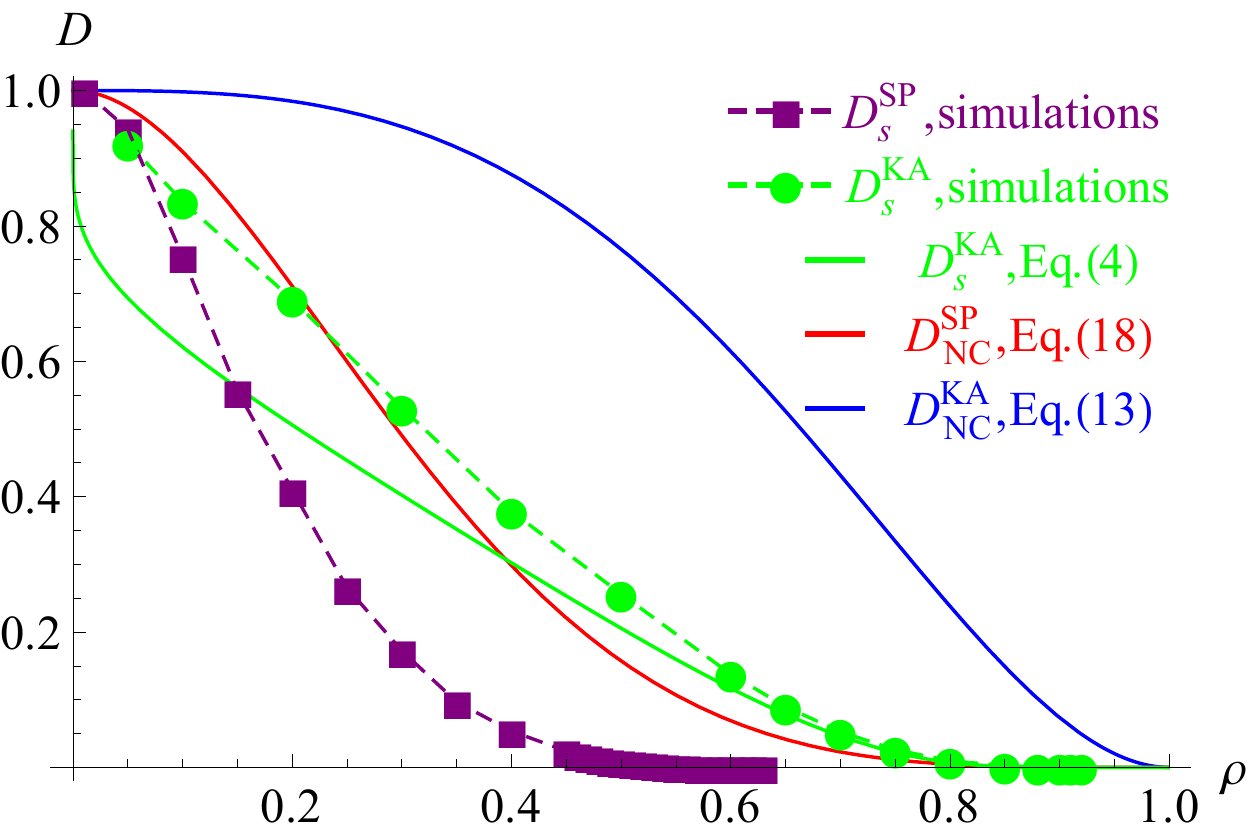}
\par\end{centering}
\caption{A comparison between the bulk diffusion coefficient, $D$, and the self-diffusion coefficient, $D_{s}$ in the $KA$ $m=2$ model and in the spiral model (SP). Symbols are results of numerical simulations. The analytical approximation for $D^{KA}_{s}$, Eq. (\ref{dska}), agrees with the numerical simulations only for $\rho\geq0.6$. The dashed lines for the self-diffusion coefficients are a guide to the eye. The results for the bulk diffusion are our no-correlation (NC) approximation, Eq. (\ref{diffka2}) for the KA model and Eq. (\ref{diffsp}) for the spiral model.}
\label{diffcompare}
\end{figure}

Note that in this paper we focus on bulk diffusion, which in general is different from self-diffusion (see Fig. \ref{diffcompare}) that is defined from the mean squared displacement in equilibrium
\begin{align}
D_{self}=\frac{1}{2d}\lim_{t\rightarrow\infty}\frac{\left\langle r^{2}\right\rangle}{t} ,
\end{align}
where $r$ is the distance after time $t$ of a particle from its initial position.
For example, in the SSEP model, the bulk diffusion coefficient does not depend on the density, while the self diffusion coefficient decreases monotonically with increasing density~\cite{124Nakazato1980,125Arita2014}. 

In the $m=2$ KA model in two dimensions, it was shown that for high densities the self-diffusion coefficient may be approximated by~\cite{116Toninelli2003}
\begin{align}
D^{KA}_{s}\approx\exp\left(\frac{2\lambda}{\ln\rho}\right) ,\label{dska}
\end{align}
where $\lambda=\pi^{2}/18\approx0.55$ \cite{073Holroyd2003}. We note here that this value of $\lambda$ is usually associated with the joint limits of infinite system size and unity particle density, while for finite size systems with finite particle density there is an effective value of $\lambda_{eff}\approx0.25$ \cite{072Teomy2014a}.

In the two dimensional spiral model the self diffusion coefficient vanishes at the critical density of directed percolation $\rho_{c}\approx0.7$ \cite{085Ghosh2014}. We will show that in either model the bulk diffusion coefficient, however, does not vanish in an infinite system at any finite density, but it may vanish in a finite system depending on its size and on the boundary condition. Consider a finite (but large) rectangular two-dimensional system connected to two different particle reservoirs in the $x$ direction, and with periodic boundary conditions in the $y$ direction, as sketched in Fig. \ref{diffusionsketch}. The occupancy of the sites in the reservoirs are delta-correlated in both space and time, which means that at each time step a site in the left reservoir, for example, is occupied with probability $\rho_{L}$ and vacant with probability $1-\rho_{L}$ irrespective of the occupancy of any other site (in the system or in the reservoir) at any time. Therefore, when checking whether there are permanently frozen clusters in the system, the sites in the reservoirs may be considered to be vacant since their occupancy fluctuates rapidly and at some point they will be vacant.

\begin{figure}
\begin{centering}
\includegraphics[width=0.8\columnwidth]{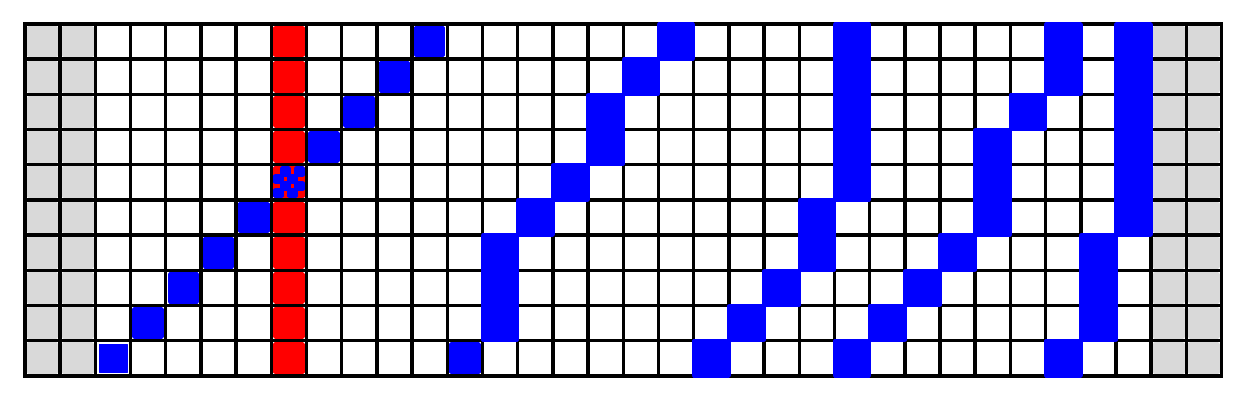}
\par\end{centering}
\caption{Frozen and non-frozen clusters in the spiral model. The gray zones are the two reservoirs. Each particle marked in red has a neighbor in its $N$ and $S$ groups (see Fig. \ref{spiral_rules} for the model's kinetic constraint), so the entire red cluster is frozen for either hard wall or periodic boundaries. For hard wall boundary conditions, each diagonal of blue particles is a different cluster, while for periodic boundary conditions all the diagonals comprise one connected cluster. Each blue particle except the two particles near the edges, has neighbors in its $N$ and $S$ groups. For hard wall boundaries, the walls act as occupied neighbors for the blue particles at the edges and so all the blue particles are frozen. For periodic boundary conditions, the blue particles at the edges are mobile and thus none of the blue particles are frozen.}
\label{spiral_diffusion_blocked}
\end{figure}

In order for the bulk diffusion to vanish, there must be frozen clusters in the system. In the spiral model, a completely occupied column acts as such a frozen cluster. The probability of at least one such fully occupied column occurring is $1-\left(1-\rho^{L_{y}}\right)^{L_{x}}$, which for $L_{x}=L_{y}=100$ is smaller than $3\times10^{-3}$ even for the highest density we consider here $(\rho=0.9)$, and is thus negligible. Another possible frozen cluster is a directed path along the diagonal. However, in order for this directed cluster to appear it must be held at its edges by other frozen particles, and in the setup we consider here the reservoirs at the $x$ direction are not frozen, see Fig. \ref{spiral_diffusion_blocked}. If there were hard-wall boundary conditions in the $y$ direction, and the system was long enough, there would have been frozen clusters spanning the system above the critical density and the bulk diffusion coefficient would vanish. In the finite systems we simulate we consider only periodic boundary conditions, so the bulk diffusion coefficient continuously goes to zero only at $\rho=1$ both for the KA model and for the spiral model.

Figure \ref{diffcompare} shows the substantial qualitative differences between the bulk diffusion coefficients and the self-diffusion coefficients in the $m=2$ KA model and in the spiral model. For the bulk diffusion coefficients, the plot shows the approximations, which we derive in the following sections, Eq. (\ref{diffka2}) for the KA model and Eq. (\ref{diffsp}) for the spiral model.

\section{No-correlations approximation for the diffusion coefficient}
\label{sec_nc}

In this section we first derive an approximation for the diffusion coefficient by neglecting correlations between the occupancy of different sites, and after that we show that these correlation are in fact important in KCMs. Nonetheless the approximate results are very close to the results of numerical simulations that we subsequently present. We define by $n_{\alpha}\left(\vec{r},t\right)$ the occupancy of site $\vec{r}$ at time $t$ under the stochastic dynamic trajectory $\alpha$, and by
\begin{align}
\rho\left(\vec{r},t\right)=\left\langle n_{\alpha}\left(\vec{r},t\right)\right\rangle
\end{align}
the occupancy of site $\vec{r}$ at time $t$ averaged over all possible stochastic trajectories. We measure time in units in which each particle attempts to move at a rate which is equal to unity, hence the evolution equation of $\rho\left(\vec{r},t\right)$ is
\begin{align}
\frac{\partial\rho\left(\vec{r},t\right)}{\partial t}=\left\langle\sum_{\hat{d}}\left[n_{\alpha}\left(\vec{r}+\hat{d},t\right)-n_{\alpha}\left(\vec{r},t\right)\right]K_{\alpha,\hat{d}}\left(\vec{r},t\right)\right\rangle ,\label{eqevdiff}
\end{align}
where $\hat{d}=\pm \hat{x}$ or $\pm \hat{y}$, and $K_{\alpha,\hat{d}}\left(\vec{r},t\right)$ encodes the kinetic-constraint such that $K=1$ if a move is possible between sites $\vec{r}$ and $\vec{r}+\hat{d}$ and $K=0$ otherwise. Note that under this construction, $K_{\alpha,\hat{d}}\left(\vec{r},t\right)$ does not depend on the occupancy of sites $\vec{r}$ and $\vec{r}+\hat{d}$. Also note that each term inside the sum in the right hand side of Eq. (\ref{eqevdiff}) is the current in the $\hat{d}$ direction between $\vec{r}+\hat{d}$ and $\vec{r}$, and the sum of all terms is the discrete divergence of the current.

Also note that in general
\begin{align}
K_{\alpha,\hat{d}}\left(\vec{r},t\right)=K_{\alpha,-\hat{d}}\left(\vec{r}+\hat{d},t\right) ,
\end{align}
and thus Eq. (\ref{eqevdiff}) may be written as
\begin{align}
&\frac{\partial\rho\left(\vec{r},t\right)}{\partial t}=\left\langle\left[n_{\alpha}\left(\vec{r}+\hat{x},t\right)-n_{\alpha}\left(\vec{r},t\right)\right]K_{\alpha,\hat{x}}\left(\vec{r},t\right)+\right.\nonumber\\
&+\left[n_{\alpha}\left(\vec{r}-\hat{x},t\right)-n_{\alpha}\left(\vec{r},t\right)\right]K_{\alpha,\hat{x}}\left(\vec{r}-\hat{x},t\right)+\nonumber\\
&+\left[n_{\alpha}\left(\vec{r}+\hat{y},t\right)-n_{\alpha}\left(\vec{r},t\right)\right]K_{\alpha,\hat{y}}\left(\vec{r},t\right)+\nonumber\\
&\left.+\left[n_{\alpha}\left(\vec{r}-\hat{y},t\right)-n_{\alpha}\left(\vec{r},t\right)\right]K_{\alpha,\hat{y}}\left(\vec{r}-\hat{y},t\right)\right\rangle .\label{eq2}
\end{align}
From the translational symmetry in the $y$ direction, we find that the last two terms cancel each other, and thus at the steady state the current
\begin{align}
J_{\hat{x}}=\left\langle\left[n_{\alpha}\left(\vec{r}+\hat{x},t\right)-n_{\alpha}\left(\vec{r},t\right)\right]K_{\alpha,\hat{x}}\left(\vec{r},t\right)\right\rangle ,
\end{align}
does not depend on $\vec{r}$, as expected.

\subsection{Kob-Andersen model}
In the $m=2$ KA model for a particle to move it needs to have at least two neighboring vacancies both before and after the move. On the square lattice, since initially the target site is vacant and finally the origin site is vacant, this is equivalent to requiring that not all the three remaining sites are occupied. We may write this condition as
\begin{align}
&K^{KA}_{\alpha,\hat{d}}\left(\vec{r},t\right)=\left[1-n_{\alpha}\left(\vec{r}-\hat{d}\right)n_{\alpha}\left(\vec{r}+\hat{d}^{\perp}\right)n_{\alpha}\left(\vec{r}-\hat{d}^{\perp}\right)\right]\nonumber\\
&\left[1-n_{\alpha}\left(\vec{r}+2\hat{d}\right)n_{\alpha}\left(\vec{r}+\hat{d}+\hat{d}^{\perp}\right)n_{\alpha}\left(\vec{r}+\hat{d}-\hat{d}^{\perp}\right)\right] ,
\end{align}
where $\hat{d}^{\perp}$ is the perpendicular direction to $\hat{d}$, and we dropped the dependence of $n_{\alpha}$ on $t$ for brevity. 

We now introduce an uncontrolled approximation in which we neglect correlations between occupancies of sites, such that for any group of sites $G$
\begin{align}
\left\langle\prod_{\vec{r}\in G}n_{\alpha}\left(\vec{r},t\right)\right\rangle=\prod_{\vec{r}\in G}\left\langle n_{\alpha}\left(\vec{r},t\right)\right\rangle=\prod_{\vec{r}\in G}\rho\left(\vec{r},t\right) ,
\end{align}
and consider the limit $L\rightarrow\infty$. In this limit the gradients are weak, and we may therefore expand Eq. (\ref{eqevdiff}) to second order in the gradients around $\vec{r}$. We then obtain Eq. (\ref{diffeq}) with the diffusion coefficient given by
\begin{align}
D_{NC}\left(\rho\right)=\overline{K}\left(\rho\right) ,\label{dnc}
\end{align}
where the subscript $NC$ represents the fact that we assumed there are no correlations, and $\overline{K}\left(\rho\right)$ is the function $K$ with each $n_{\alpha}\left(\vec{r},t\right)$ replaced by $\rho$. For example, in the $m=2$ KA model in two dimensions we find
\begin{align}
D_{NC}^{KA}(\rho)=\overline{K}^{KA}\left(\rho\right)=\left(1-\rho^{3}\right)^{2} .\label{diffka2}
\end{align}
This results from the fact that the kinetic constraint requires that at least one of the three neighbors (not including the target site) of the origin site is empty and at least one of the corresponding three sites of the target site is empty. 

This expression for the diffusion coefficient is very similar to the NC approximation of the current in the presence of an external field \cite{Sellitto2008,130Turci2012a}. In those papers, the authors considered the $m=2$ KA model on a square two-dimensional lattice with periodic boundary conditions on all sides, but with a homogeneous applied field in the $x$ direction, such that a particle moves in the negative $x$ direction with a lower probability than in the other three directions. In the extreme case in which the particle cannot move against the field, the average current is given by
\begin{align}
J^{field}_{\vec{r},d}=\left\langle n_{\alpha}\left(\vec{r},t\right)\left[1-n_{\alpha}\left(\vec{r}+\hat{x},t\right)\right]K_{\alpha,\hat{x}}\left(\vec{r},t\right)\right\rangle ,
\end{align}
which in the steady state under the no correlations approximation may be expressed using the diffusion coefficient
\begin{align}
J^{field}_{NC}\left(\rho\right)=\rho\left(1-\rho\right)D_{NC}\left(\rho\right) .
\end{align}
Similarly to what we find here, this approximation works rather well at low densities and exhibits small deviations at high densities~\cite{Sellitto2008,130Turci2012a}.

Note that in general $\overline{K}\left(\rho\right)$ is equivalent to the probability that a move is possible between two adjacent sites given that one of them is occupied and the other is vacant. In the KA model in general dimensions and $m\leq d$, a particle can move if at least $m$ of its neighbors are vacant before and after the move. Since the target site is vacant before the move and the origin site is vacant after the move, the kinetic rule is equivalent to saying that a particle can move to an adjacent vacant site if at least $m-1$ of its neighbors, not including the target site, are vacant, and that at least $m-1$ of the neighbors of the target site, not including the origin site, are vacant. Since there is no overlap between the $2d-1$ neighbors of the origin site (excluding the target site), and between the $2d-1$ neighbors of the target site (excluding the origin site), we find that
\begin{align}
&D_{NC}^{KA}(\rho)=\overline{K}^{KA}\left(\rho\right)=\nonumber\\
&=\left[1-\sum^{m-2}_{n=0}\left(\begin{array}{c}2d-1\\n\end{array}\right)\left(1-\rho\right)^{n}\rho^{2d-1-n}\right]^{2} ,
\end{align}
where the summation variable $n$ is the number of nearest neighbor vacant sites. Performing the sum yields
\begin{align}
&D_{NC}^{KA}(\rho)=\overline{K}^{KA}\left(\rho\right)=\nonumber\\
&=\left(m-1\right)^{2}B^{2}\left(-\frac{1-\rho}{\rho};m-1,1-2d\right) ,\label{kadiffgen}
\end{align}
where $B\left(z;a,b\right)$ is the incomplete beta function~\cite{128WeissteinIncompleteBeta}. 

\subsection{Spiral Model}
\begin{figure}
\begin{centering}
\includegraphics[width=0.8\columnwidth]{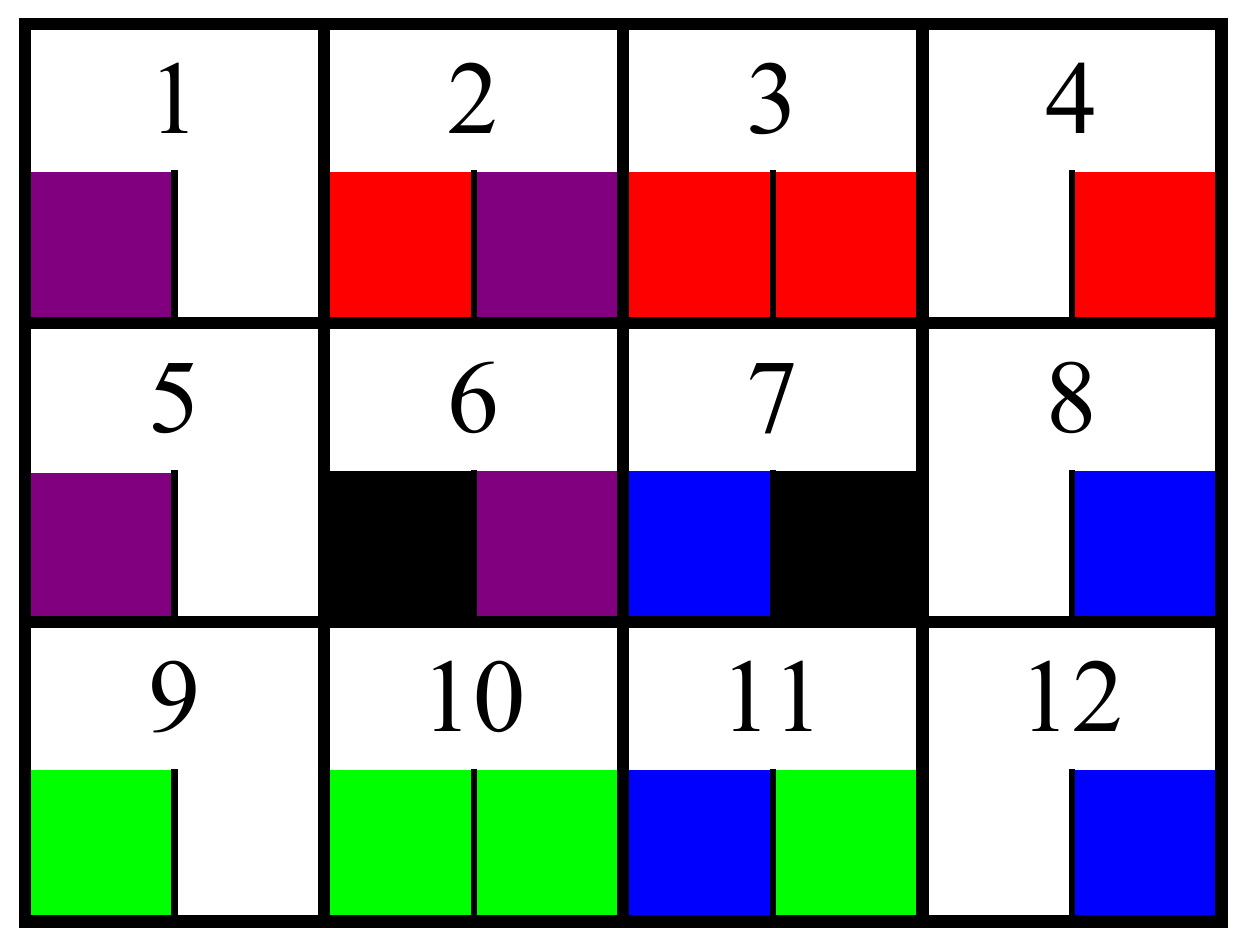}
\par\end{centering}
\caption{A graphic illustration of the rules for an attempted move between sites $6$ and $7$ in the spiral model. At each site of the twelve sites, the color on the left corresponds to its relation to site $6$ and the color on the right corresponds to its relation to site $7$. The four groups are defined by color: $N$ (red), $S$ (green), $E$ (blue) and $W$ (purple). The black color represents the origin/target sites.}
\label{spiraldiff}
\end{figure}

For the two-dimensional spiral model, the kinetic constraint involves the ten sites surrounding the origin and target sites. We follow the labeling of sites in Fig. \ref{spiraldiff} and assume that a particle attempts to move from site $6$ to site $7$, such that site $6$ is occupied before the move and site $7$ is vacant before the move. In order for the particle to move (see Fig. \ref{spiral_rules} above) it needs that before the move either both its $N$ neighbors (sites $2$ and $3$) are vacant or both its $S$ neighbors (sites $9$ and $10$) are vacant, and that either both its $E$ neighbors (sites $7$ and $11$) are vacant or both its $W$ neighbors (sites $1$ and $5$) are vacant. It similarly needs to obey the rule after moving, i.e. in relation to site $7$. Note that before the move site $7$ is vacant, and after the move site $6$ is vacant. In principle we now need to consider all $2^{10}$ possibilities for the occupancy of the ten surrounding sites, with each site occupied independently with probability $\rho$ and vacant with probability $v=1-\rho$. However, we first note that sites $2$, $3$, $10$ and $11$ are neighbors of both sites $6$ and $7$, so we start checking from them.
\begin{itemize}
\item Case $1$: If all four sites $2$, $3$, $10$ and $11$ are vacant, then the $W$ and $S$ groups of site $7$ and the $N$ and $E$ groups of site $6$ are all vacant, which means that the move is allowed. This occurs with probability $\left(1-\rho\right)^{4}$.
\item Case $2$: If site $3$ is occupied and sites $2$, $10$ and $11$ are vacant, then the $W$ and $S$ groups of site $7$ and the $E$ group of site $6$ are all vacant. The $N$ group of site $6$ is not all vacant, so in order for the move to be allowed, its $S$ group must be all vacant, i.e. site $9$ must be vacant. This occurs with probability $\left(1-\rho\right)^{4}\rho$. From symmetry, this is equivalent to the case in which site $10$ is occupied and sites $2$, $3$ and $11$ are vacant.
\item Case $3$: If site $2$ is occupied and sites $10$ and $11$ are vacant (regardless of the state of site $3$), then the $S$ group of site $7$ and the $E$ group of site $6$ are all vacant, while the $W$ group of site $7$ and the $N$ group of site $6$ are not all vacant. Thus, in order to facilitate the move, sites $8$, $9$ and $12$ must also be vacant. This occurs with probability $\left(1-\rho\right)^{5}\rho$. From symmetry, this is equivalent to the case in which site $11$ is occupied and sites $2$ and $3$ are vacant.
\item Case $4$: If sites $2$ and $11$ are occupied, then the $W$ and $S$ groups of site $7$ and the $N$ and $E$ groups of site $6$ are not all vacant, which means that sites $1$, $3$, $4$, $5$, $8$, $9$, $10$ and $12$ must be vacant. This occurs with probability $\left(1-\rho\right)^{8}\rho^{2}$.
\end{itemize} 
In all other cases the move is not allowed. Therefore, summing all the cases, we find that the probability that the move is allowed is given by
\begin{align}
&D_{NC}^{SP}(\rho)=\overline{K}^{SP}\left(\rho\right)=\nonumber\\
&=\left(1-\rho\right)^{4}+2\left(1-\rho\right)^{4}\rho+2\left(1-\rho\right)^{5}\rho+\left(1-\rho\right)^{8}\rho^{2}=\nonumber\\
&=\left(1-\rho\right)^{4}\left[1+2\rho\left(2-\rho\right)+\left(1-\rho\right)^{4}\rho^{2}\right] ,\label{diffsp}
\end{align}
which constitutes the no-correlation approximation for the diffusion coefficient in the spiral model.

\section{Density profiles}
\label{secdenprof}
\begin{figure}
\begin{centering}
\includegraphics[width=0.95\columnwidth]{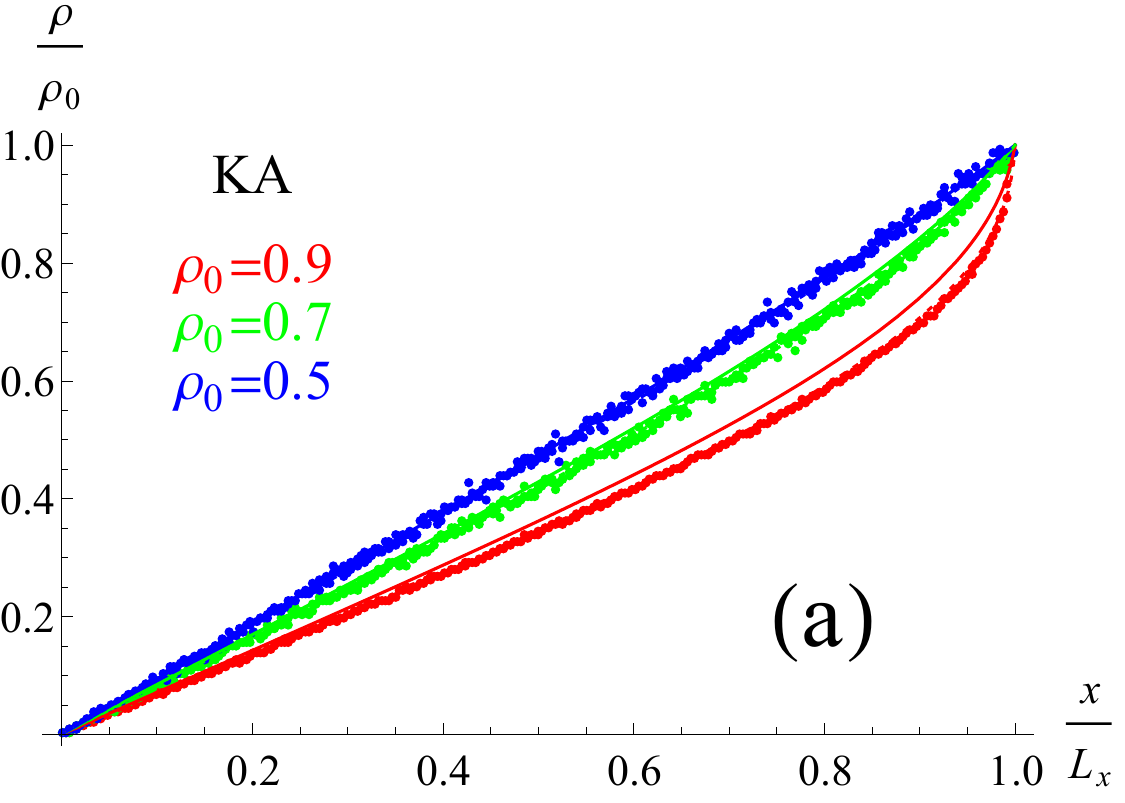}\\
\includegraphics[width=0.95\columnwidth]{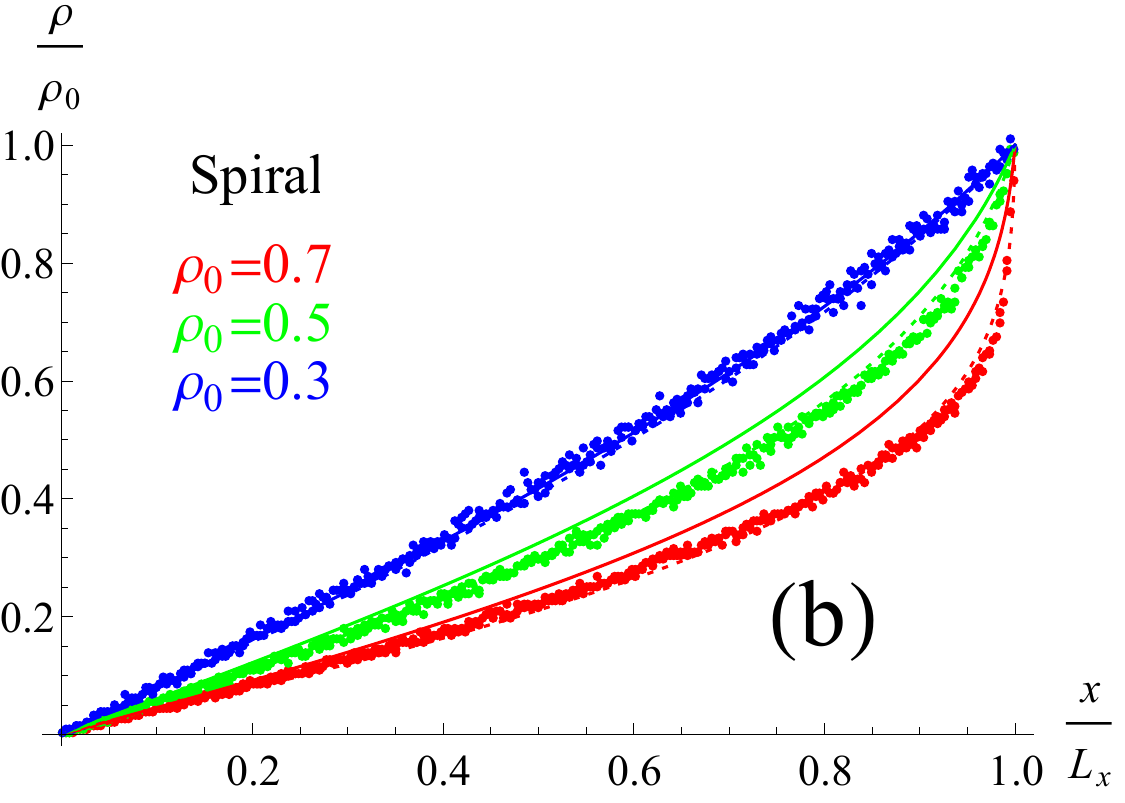}\\
\par\end{centering}
\caption{The normalized density profiles at the steady state for the KA model (a) and the spiral model (b). The symbols are from the simulations done on a $400\times100$ system, the solid lines are the NC approximation, Eq. (\ref{mhd}), and the dashed lines are the NC$+\rho_{eff}$ approximation, Eq. (\ref{deff}). Some of the lines cannot be seen since they fall on the simulations.}
\label{mdif1}
\end{figure}

We now want to check the quality of the NC approximation. We do this by comparing the density profile $\rho_{NC}(x,t)$ found by solving the nonlinear diffusion equation, Eq. (\ref{diffeq}), with $D(\rho)=D_{NC}(\rho)$, Eqs. (\ref{diffka2}) and (\ref{diffsp}), and the density profile $\rho(x,t)$ obtained from numerical simulations of the two models. Figure \ref{mdif1} shows that the steady state density profiles in the simulations are lower than in the approximation. This means that the correlations between sites, which we neglected, cause the true diffusion coefficient to be lower than its approximation. We will show that the difference between the two is not merely a finite size effect, but a genuine difference. The setup we consider is that the initial condition in both cases is $\rho(x,0)=\rho_{0}$, and the density of the reservoir at $x=0$ is $\rho_{L}=0$, and at $x=L_{x}$ is $\rho_{R}=\rho_{0}$. In order to compare different system sizes we plot the density as a function of the normalized position $x/L_x$.

\begin{figure}
\begin{centering}
\includegraphics[width=0.95\columnwidth]{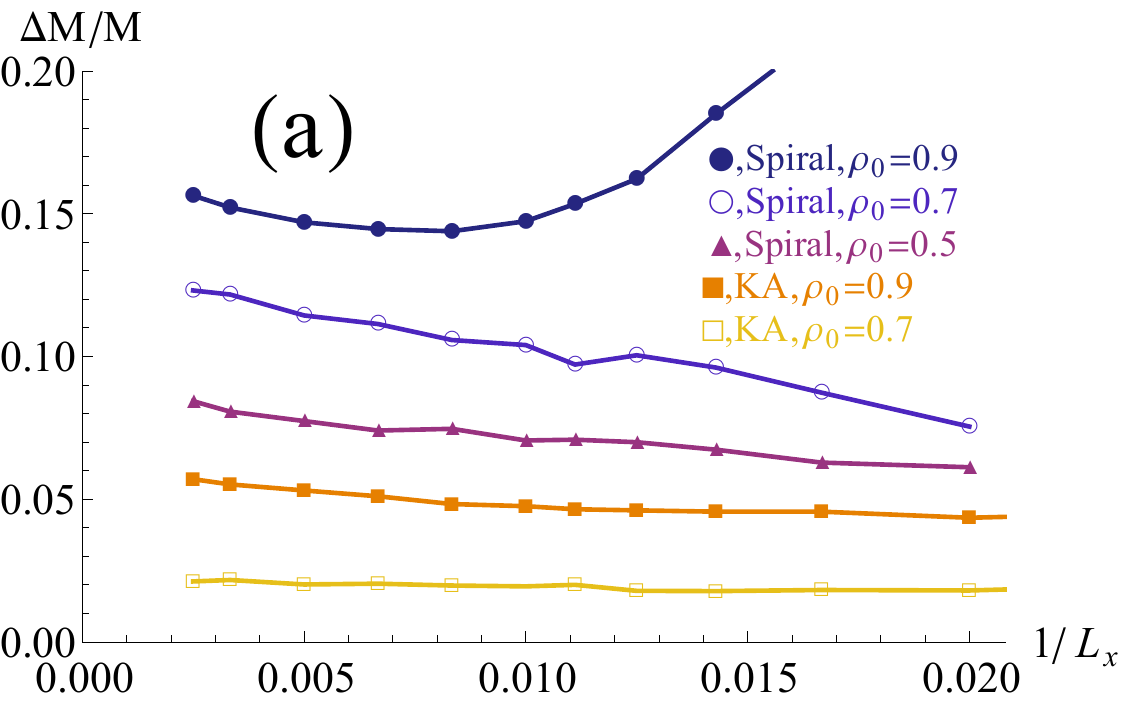}\\
\includegraphics[width=0.95\columnwidth]{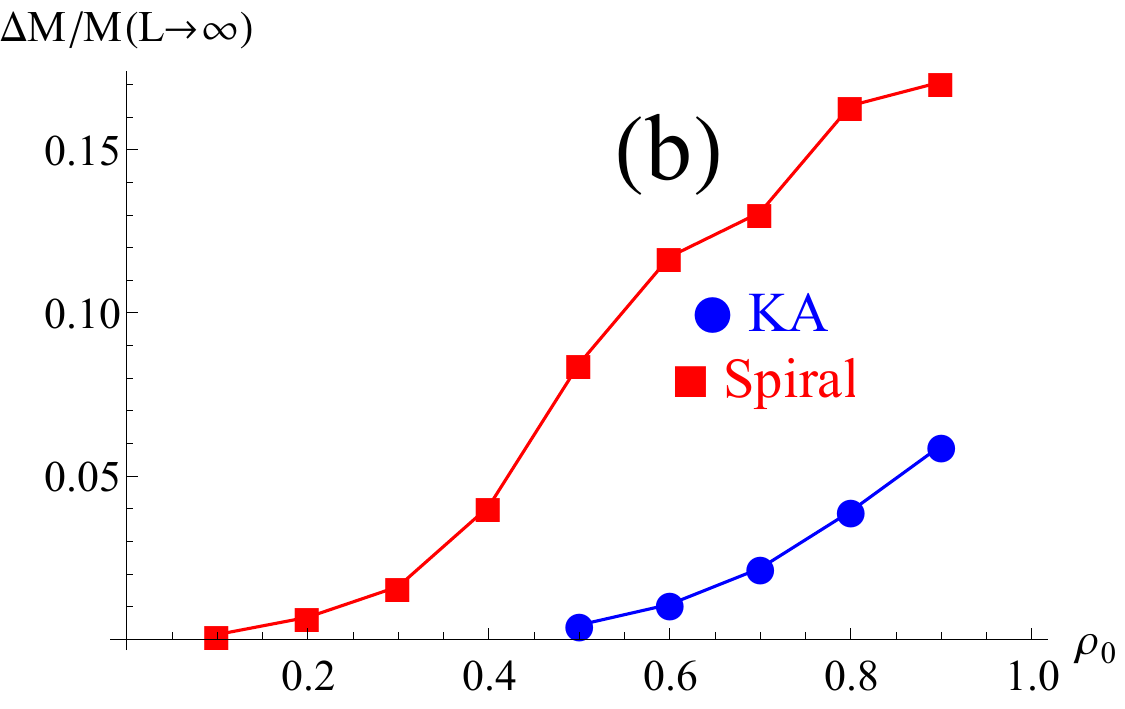}
\par\end{centering}
\caption{(a) The relative difference between the total mass in the steady state according to the NC approximation and between the simulations, $\frac{\Delta M}{M_{NC}}$. (b) An extrapolation of $\Delta M/M_{NC}$ at $L\rightarrow\infty$. No data is shown for the KA model at $\rho_{0}\leq0.4$ since the difference there is very hard to resolve and is practically equal to zero.}
\label{mdif2}
\end{figure}

To quantify the difference between the numerical result and our NC approximation we compare the total normalized mass in the system at the steady state
\begin{align}
M=\frac{1}{L_{x}}\int^{L_{x}}_{0}\rho(x)dx .\label{mdef}
\end{align}
In the steady state, we integrate the diffusion equation with respect to $x$ such that
\begin{align}
c_{1}=D\left[\rho(x)\right]\frac{d\rho}{dx} .\label{1int}
\end{align}
Integrating again with respect to $x$ yields
\begin{align}
c_{1}x+c_{0}=\int^{\rho(x)}_{\rho(0)}D\left[\rho\left(x'\right)\right]\frac{d\rho(x')}{dx'}dx' .\label{2int}
\end{align}
Setting $x=0$ in Eq. (\ref{2int}) yields $c_{0}=0$. Setting $x=L_{x}$ in Eq. (\ref{2int}) yields
\begin{align}
c_{1}L_{x}=\int^{\rho_{R}}_{\rho_{L}}D\left(\rho\right)d\rho .\label{c1def}
\end{align}
Combining Eqs. (\ref{mdef}) ,(\ref{1int}) and (\ref{c1def}) yields
\begin{align}
M=\frac{\int^{\rho_{R}}_{\rho_{L}}\rho D(\rho)d\rho}{\int^{\rho_{R}}_{\rho_{L}}D(\rho)d\rho} .\label{mhd}
\end{align}
Therefore, we can analytically compute $M_{NC}$ for the above polynomial expressions, Eqs. (\ref{diffka2}) and (\ref{diffsp}), for the no-correlations approximate diffusion coefficient $D_{NC}\left(\rho\right)$.

From Fig. \ref{mdif2}a we see that $\Delta M=M_{NC}-M\left(L_{x}\right)$ does not converge to $0$ as $L_{x}$ is increased, which means that the difference is clearly not a finite-size effect. From Fig. \ref{mdif2}b we see that the difference grows with the density, as expected.

One may argue that the difference arises due to the large density gradients in the system which are not well resolved in the simulation. However, in the steady state, the largest gradient in the density profile, $G_{max}=L_{x}max\left[\rho\left(x\right)-\rho\left(x-1\right)\right]$, is at $x=L_{x}$, and in the cases we checked, we find numerically that $G^{KA}_{max}\left(\rho_{0}=0.8\right)\approx2.6$, $G^{KA}_{max}\left(\rho_{0}=0.9\right)\approx8.7$, $G^{SP}_{max}\left(\rho_{0}=0.5\right)\approx1.9$, $G^{SP}_{max}\left(\rho_{0}=0.7\right)\approx14$, and $G^{SP}_{max}\left(\rho_{0}=0.9\right)\approx1060$. To resolve this gradient we require $L_{x}\gg G_{max}$, which is satisfied in all the cases we checked except for the spiral model at $\rho_{0}=0.9$. We further note that while increasing $L_{x}$ adds data points close to $x=L_{x}$, it does not change the density profile evaluated at smaller $x/L_{x}$. We verified that in the simulations the system indeed reached the steady state by first numerically solving the diffusion equation with $D=D_{NC}$ and finding the time it takes the total mass of the system to reach within $1\%$ of its value in the steady state, and then running the dynamical simulations up to a time which is ten times longer than that. 

From these results we see that even in the most extreme case, the relative difference in the density profile between the NC approximation and the simulations is at most $0.2$. This is rather good as an approximation, however it is much larger than the relative difference seen in other non-gradient lattice models using the same type of approximation~\cite{125Arita2014}. In particular, KCMs are in general non-gradient models and thus we expect the diffusion coefficient to differ from the NC approximation.

\begin{figure}
\begin{centering}
\includegraphics[width=0.475\columnwidth]{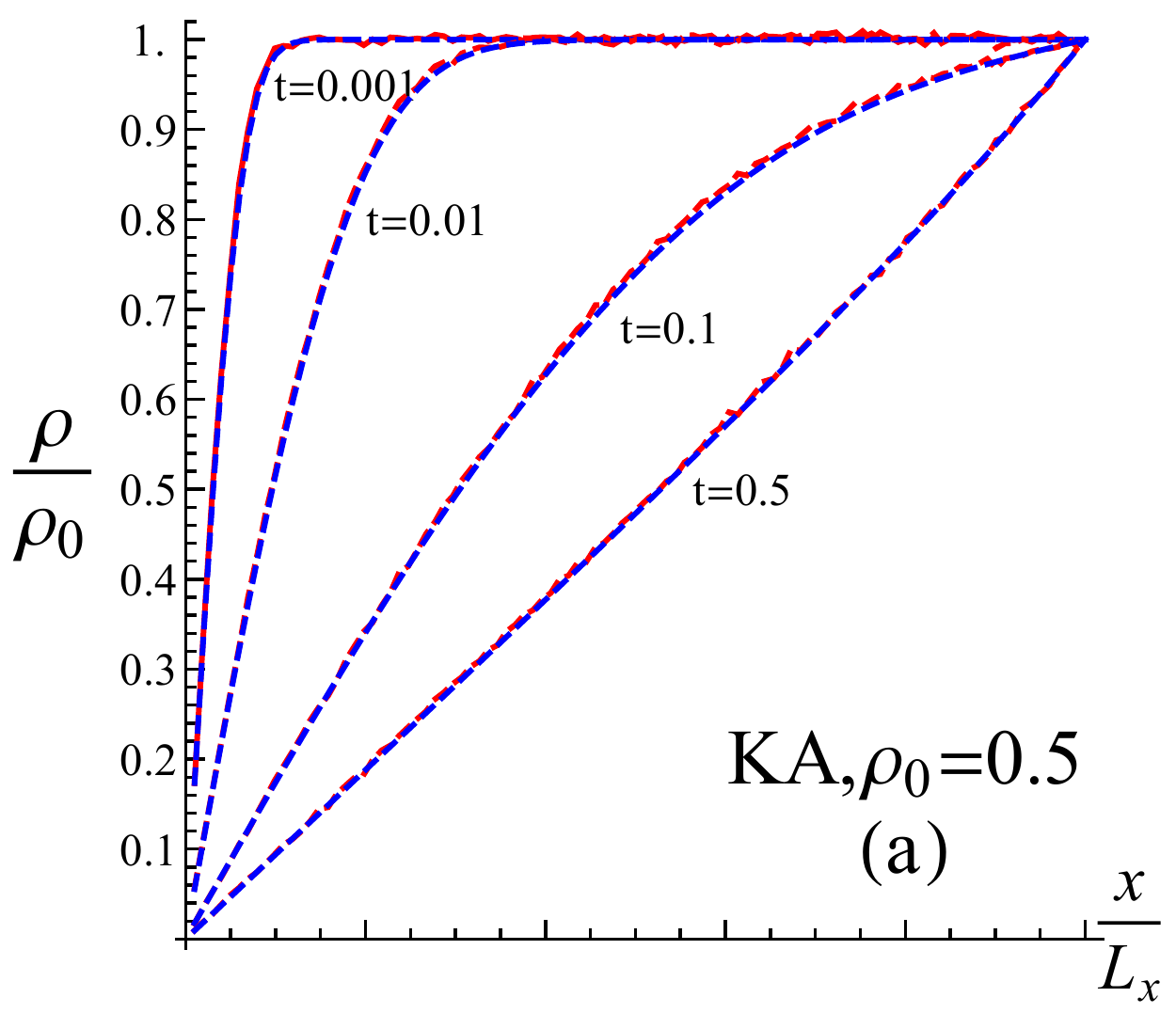}
\includegraphics[width=0.475\columnwidth]{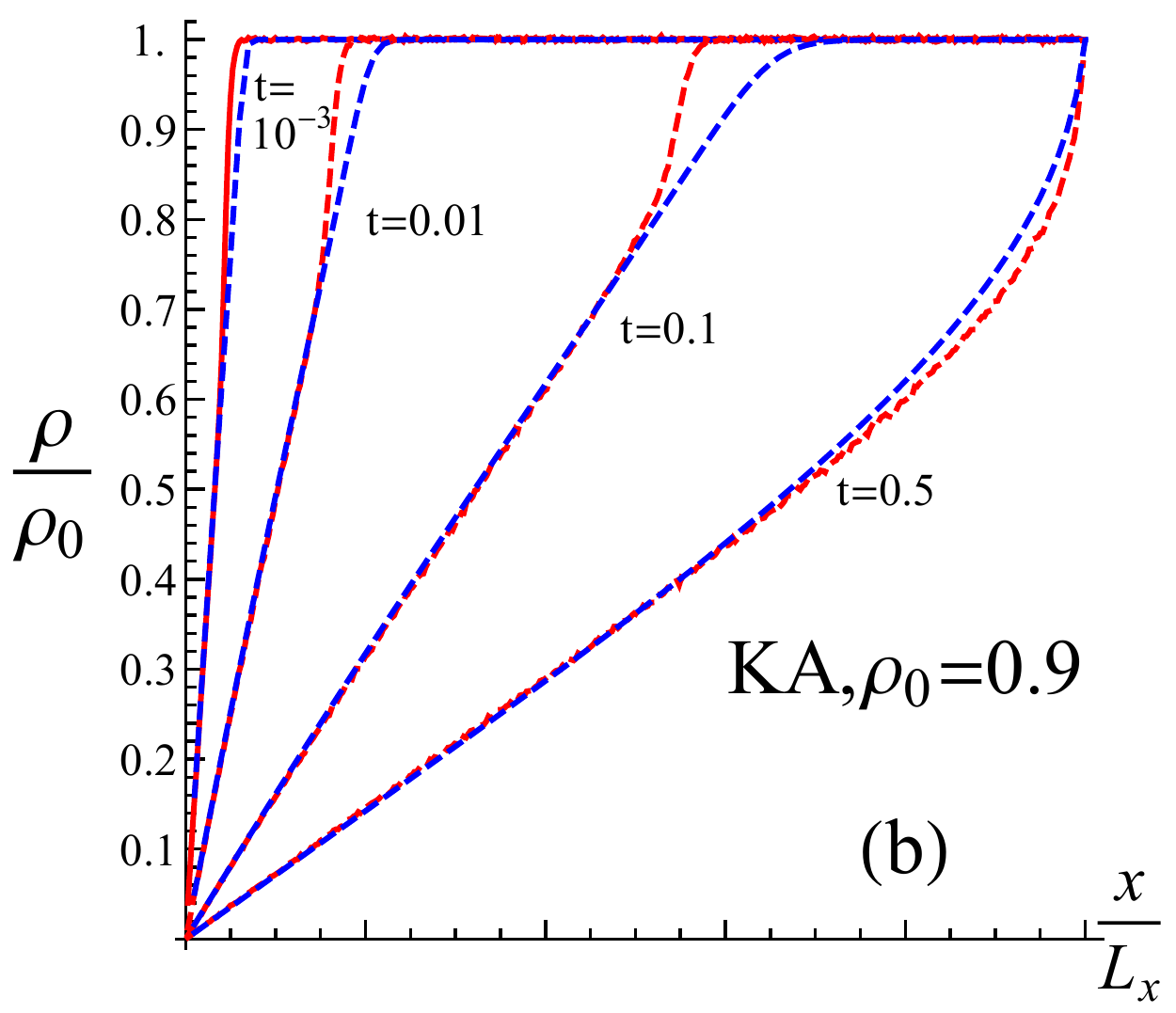}\\
\includegraphics[width=0.475\columnwidth]{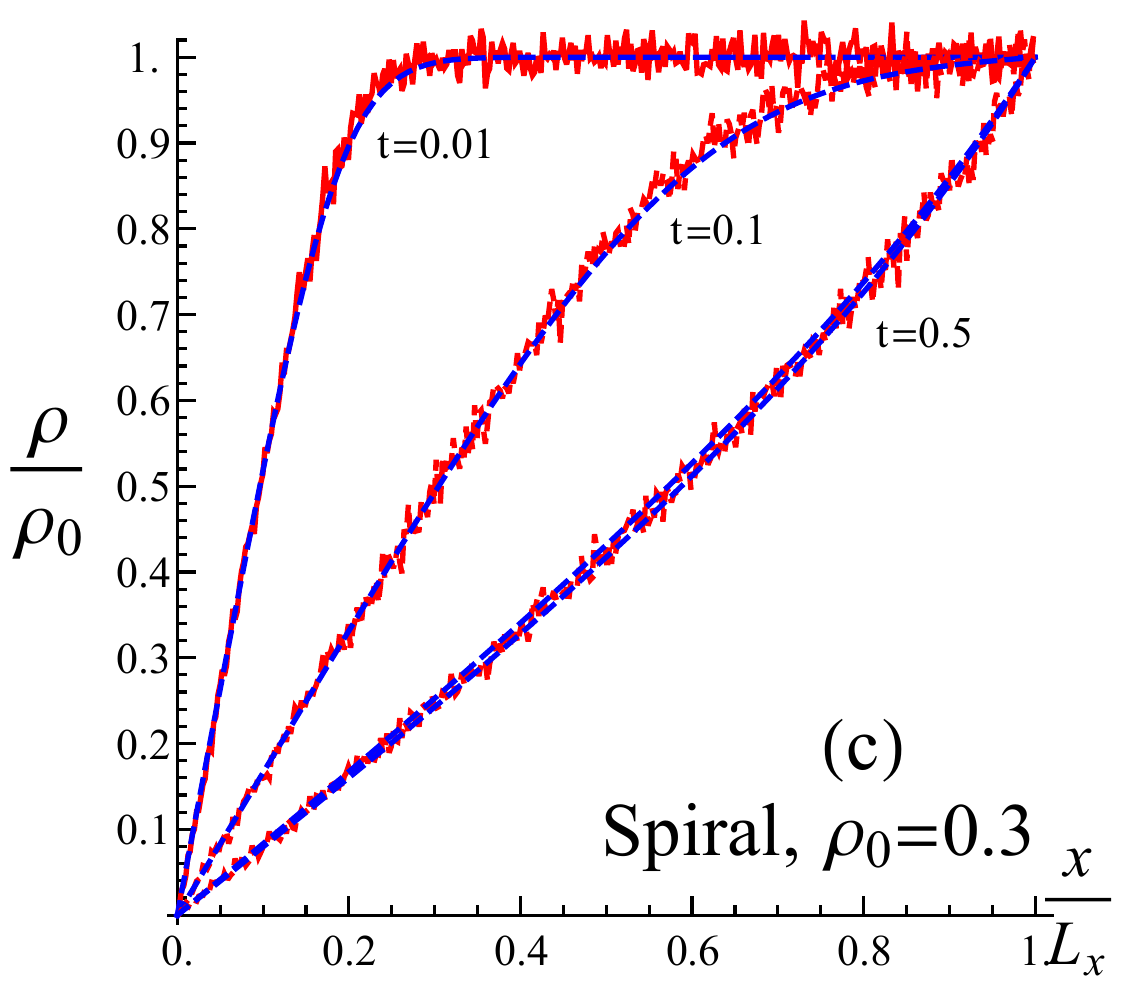}
\includegraphics[width=0.475\columnwidth]{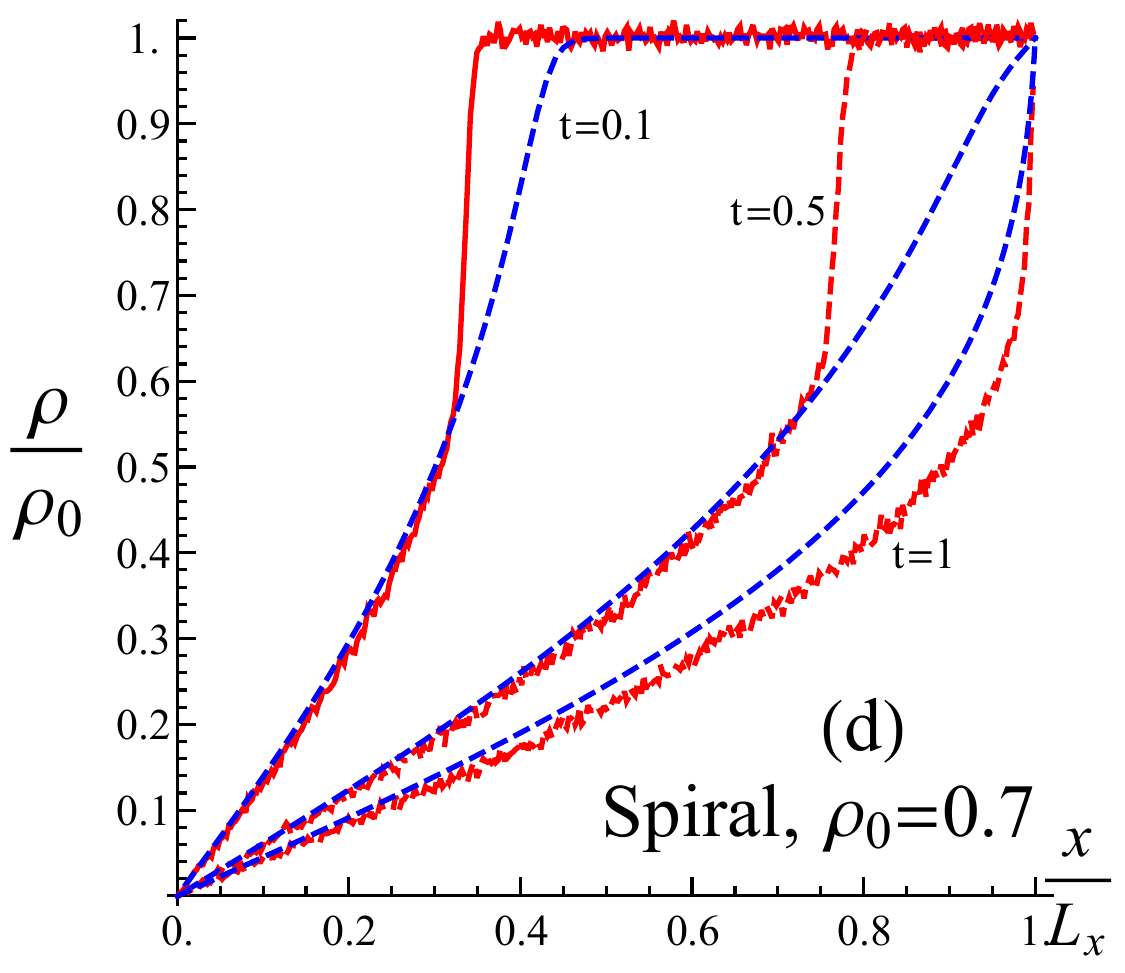}\\
\par\end{centering}
\caption{The density profile from the NC approximation (dashed blue) and the $400\times100$ simulations (red) at different times for the KA model and the spiral model at different densities.}
\label{vdif1}
\end{figure}

\begin{figure}
\begin{centering}
\includegraphics[width=0.8\columnwidth]{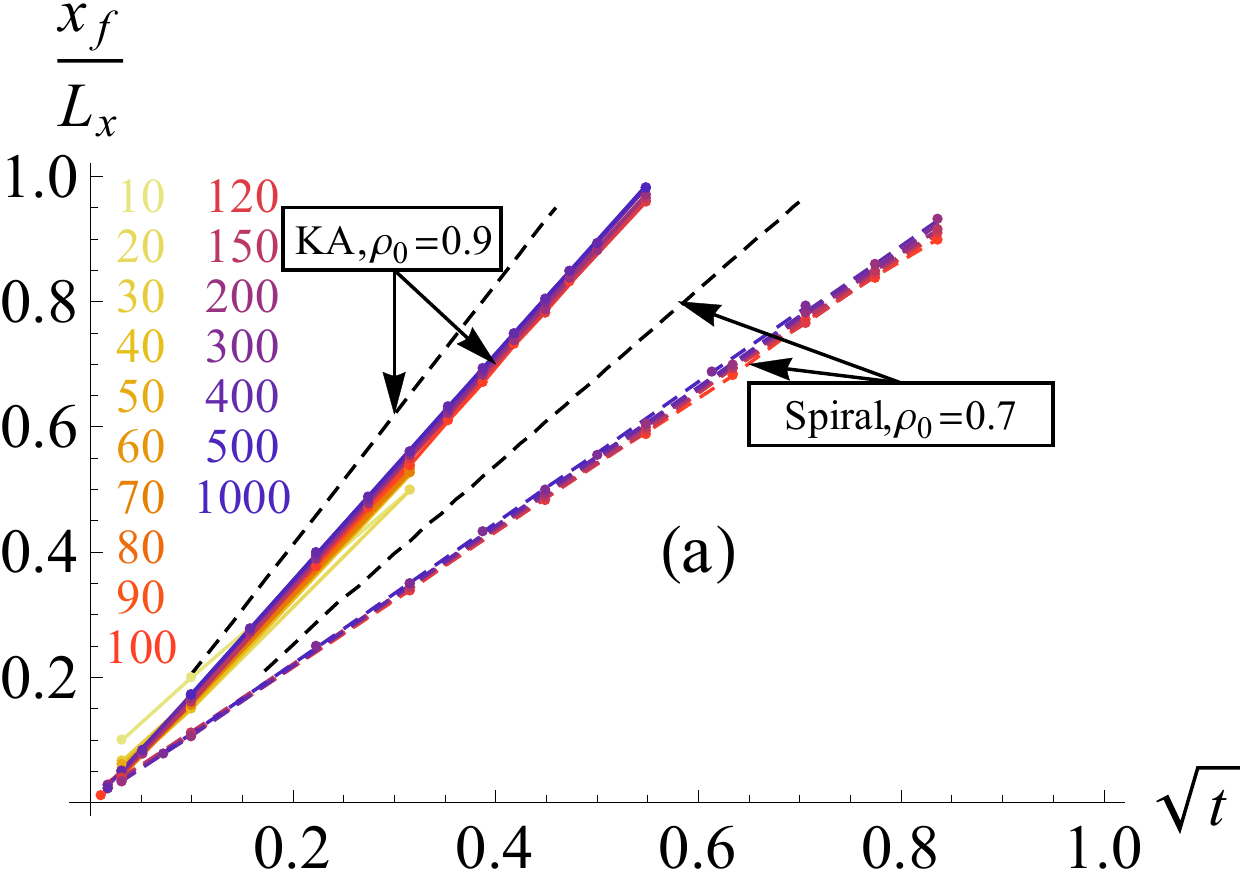}\\
\includegraphics[width=0.8\columnwidth]{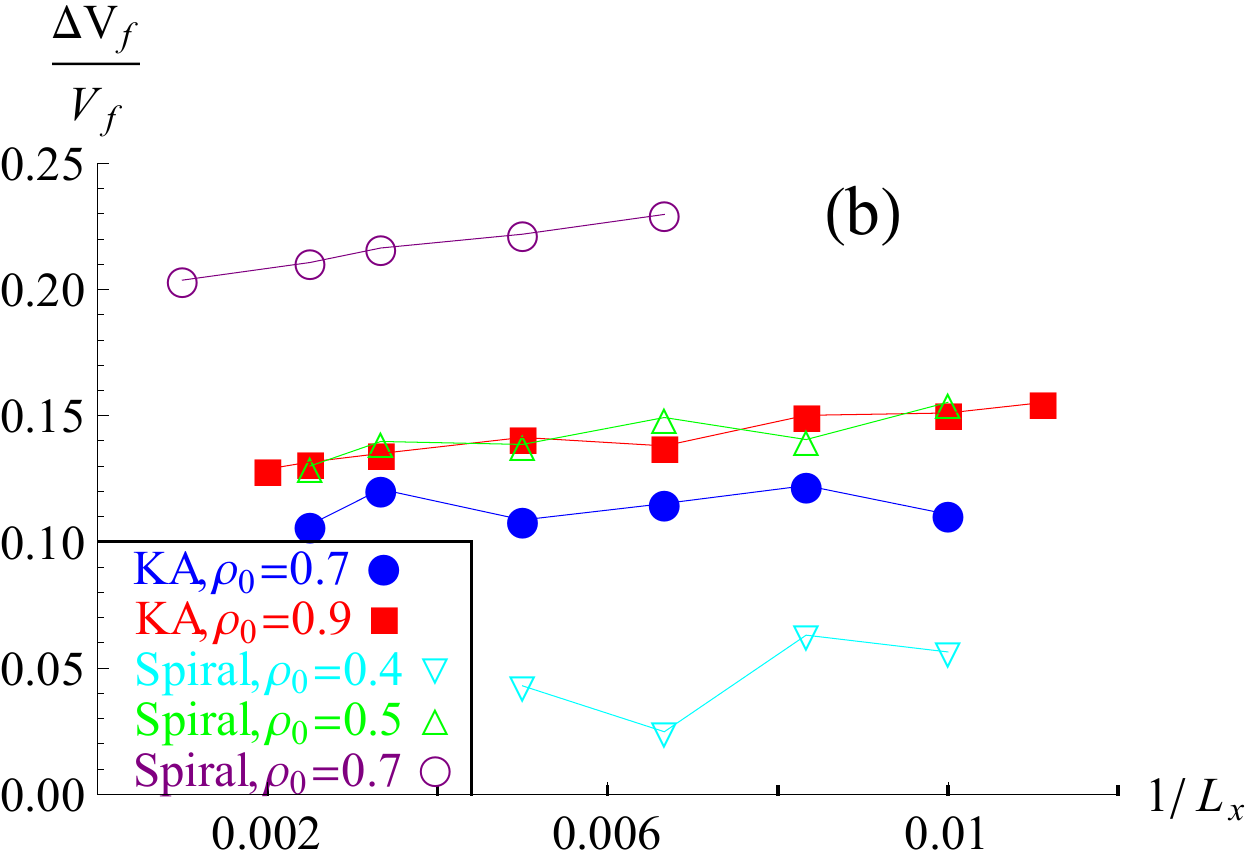}\\
\par\end{centering}
\caption{(a) The propagation of the front for the spiral model at $\rho_{0}=0.7$ and the KA model at $\rho_{0}=0.9$ for various system sizes as indicated in the legend. The dashed line is the NC solution which substantially deviates from the simulation results. (b) The relative difference of $V_{f}$ between the simulations and the NC solution as a function of system size for the two models and for various densities as indicated in the legend.}
\label{vdif2}
\end{figure}

Figure \ref{vdif1} shows the density profiles at different times. The main difference between the approximate NC solution and the simulations is a kink that appears at intermediate times before the system reaches the steady state. We investigate the dynamics of this kink by defining its position as the value of $x$ for which $\rho(x_{f},t)=\rho_{0}-\delta$, where we arbitrarily choose $\delta=0.02$. From Fig. \ref{vdif2}a we see that
\begin{align}
\frac{x_{f}}{L_{x}}=V_{f}\sqrt{t} ,
\end{align}
in both the simulations and the NC solution. The scaling of $x_{f}$ with $\sqrt{t}$ is expected as this is a diffusive process. However, while $V_{f}$ from the simulations appears roughly independent of system size, it is substantially different from $V_{f}$ from the NC solution. Again, from Fig. \ref{vdif2}b we see that it is not a finite-size effect, but a genuine difference.

\section{Correlations}
\label{sec_corr}

If $\rho_{L}=\rho_{R}$ the system is in equilibrium and there are indeed no correlations between sites. However, when $\rho_{L}\neq\rho_{R}$ there are correlations. The question is whether they have an effect on the diffusion coefficient. We will show that in gradient models the correlations do not affect the diffusion coefficient, while in non-gradient models they might affect it.

\begin{figure}
\begin{centering}
\includegraphics[width=0.95\columnwidth]{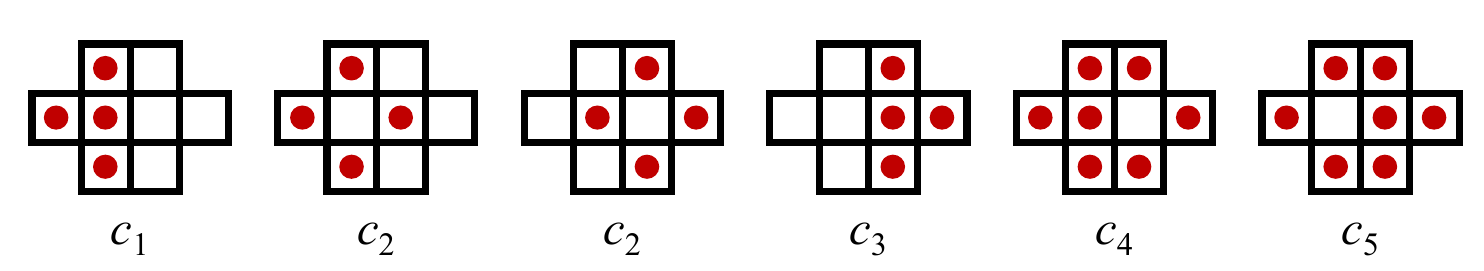}
\par\end{centering}
\caption{The six groups of sites included in the correlation function of the KA model.}
\label{corrkafig}
\end{figure}

Consider a very small density gradient such that
\begin{align}
\rho_{R}=\rho_{L}+\epsilon ,
\end{align}
with $\epsilon\ll1$. The small density gradient creates a small current in the system
\begin{align}
J_{\vec{r},\hat{d}}=\left\langle\left[n_{\alpha}\left(\vec{r}+\hat{d},t\right)-n_{\alpha}\left(\vec{r},t\right)\right]K_{\alpha,\hat{d}}\left(\vec{r},t\right)\right\rangle ,
\end{align}
which at the steady state is independent of $\vec{r}$. The diffusion coefficient of the system is defined by
\begin{align}
D=\lim_{\begin{array}{c}L_{x}\rightarrow\infty\\\epsilon\rightarrow0\end{array}}J_{\vec{r},\hat{x}}\frac{L_{x}}{\epsilon} ,\label{diffdef}
\end{align}
where the current is calculated at the steady state. We now define the correlation function
\begin{align}
&C_{\vec{r},\hat{d}}=\left\langle\left[n_{\alpha}\left(\vec{r}+\hat{d},t\right)-n_{\alpha}\left(\vec{r},t\right)\right]K_{\alpha,\hat{d}}\left(\vec{r},t\right)\right\rangle-\nonumber\\
&-\left[\rho\left(\vec{r}+\hat{d},t\right)-\rho\left(\vec{r},t\right)\right]\left\langle K_{\vec{r},\hat{d}}\right\rangle ,\label{corr}
\end{align}
where $\left\langle K_{\vec{r},\hat{d}}\right\rangle$ is the average taken over each site independently. As an example for non-gradient models, consider the $m=2$ KA model in two dimensions for which
\begin{align}
&\left\langle K^{KA}_{\vec{r},\hat{d}}\right\rangle=\left[1-\rho\left(\vec{r}-\hat{d},t\right)\rho\left(\vec{r}+\hat{d}^{\perp},t\right)\rho\left(\vec{r}-\hat{d}^{\perp},t\right)\right]\nonumber\\
&\left[1-\rho\left(\vec{r}+2\hat{d},t\right)\rho\left(\vec{r}+\hat{d}+\hat{d}^{\perp},t\right)\rho\left(\vec{r}+\hat{d}-\hat{d}^{\perp},t\right)\right] .\label{corrka}
\end{align}
The correlation function for the KA model is a sum of six correlation functions of groups of different sites, as shown in Fig. \ref{corrkafig}, and may be written as
\begin{align}
&C^{KA}_{\vec{r},\hat{d}}=c_{1}(\vec{r})-c_{2}(\vec{r})+c_{2}(\vec{r}+\hat{x})-c_{3}(\vec{r}+\hat{x})+\nonumber\\
&+c_{4}(\vec{r})-c_{5}(\vec{r}+\hat{x}) ,\label{corrka2}
\end{align}
with
\begin{align}
&c_{1}(\vec{r})=\left\langle n_{\alpha}(\vec{r})n_{\alpha}(\vec{r}-\hat{x})n_{\alpha}(\vec{r}+\hat{y})n_{\alpha}(\vec{r}-\hat{y})\right\rangle-\nonumber\\
&-\rho^{3}(\vec{r})\rho\left(\vec{r}-\hat{x}\right) ,\nonumber\\
&c_{2}(\vec{r})=\left\langle n_{\alpha}(\vec{r}+\hat{x})n_{\alpha}(\vec{r}-\hat{x})n_{\alpha}(\vec{r}+\hat{y})n_{\alpha}(\vec{r}-\hat{y})\right\rangle-\nonumber\\
&-\rho^{2}(\vec{r})\rho\left(\vec{r}+\hat{x}\right)\rho\left(\vec{r}-\hat{x}\right) ,\nonumber\\
&c_{3}(\vec{r})=\left\langle n_{\alpha}(\vec{r})n_{\alpha}(\vec{r}+\hat{x})n_{\alpha}(\vec{r}+\hat{y})n_{\alpha}(\vec{r}-\hat{y})\right\rangle-\nonumber\\
&-\rho^{3}(\vec{r})\rho\left(\vec{r}+\hat{x}\right) ,\nonumber\\
&c_{4}(\vec{r})=\left\langle n_{\alpha}(\vec{r})n_{\alpha}(\vec{r}-\hat{x})n_{\alpha}(\vec{r}+\hat{y})n_{\alpha}(\vec{r}-\hat{y})\right.\nonumber\\
&\left.n_{\alpha}(\vec{r}+\hat{x}+\hat{y})n_{\alpha}(\vec{r}+\hat{x}-\hat{y})n_{\alpha}(\vec{r}+2\hat{x})\right\rangle\nonumber\\
&-\rho^{3}(\vec{r})\rho\left(\vec{r}-\hat{x}\right)\rho^{2}\left(\vec{r}+\hat{x}\right)\rho\left(\vec{r}+2\hat{x}\right) ,\nonumber\\
&c_{5}(\vec{r})=\left\langle n_{\alpha}(\vec{r})n_{\alpha}(\vec{r}+\hat{x})n_{\alpha}(\vec{r}+\hat{y})n_{\alpha}(\vec{r}-\hat{y})\right.\nonumber\\
&\left.n_{\alpha}(\vec{r}-\hat{x}+\hat{y})n_{\alpha}(\vec{r}-\hat{x}-\hat{y})n_{\alpha}(\vec{r}-2\hat{x})\right\rangle\nonumber\\
&-\rho^{3}(\vec{r})\rho\left(\vec{r}+\hat{x}\right)\rho^{2}\left(\vec{r}-\hat{x}\right)\rho\left(\vec{r}-2\hat{x}\right) .
\end{align}
When the inversion symmetry is broken, as in the setup we consider here, the terms in Eq. (\ref{corrka2}) do not necessarily cancel each other. We believe that the correlations we see here are in fact strongly related to the correlations that were investigated in \cite{Sellitto2008,130Turci2012a} for the driven KA model. In equilibrium, blocked regions can become unblocked by particles moving away from these regions in all directions. When there is a preferred direction, as is the case both in our work and in \cite{Sellitto2008,130Turci2012a}, blocked regions may remain blocked for longer periods of time than in equilibrium. Since these blocked regions are more common in out of equilibrium situations than in equilibrium, there are correlations between the occupancy of neighboring sites.

The correlation function can in general be written as
\begin{align}
C_{\vec{r},\hat{d}}=C^{(1)}_{\vec{r},\hat{d}}\frac{\epsilon}{L_{x}}+o\left(\frac{\epsilon}{L_{x}}\right) ,\label{corrapp}
\end{align}
where $C^{(1)}_{\vec{r},\hat{d}}$ does not depend on $\epsilon$ or on $L_{x}$. If $C^{(1)}_{\vec{r},\hat{d}}=0$, the correlations do not affect the diffusion coefficient. We will now show that in non-gradient models it is not necessarily equal to zero in general, while in gradient models it is equal to zero.
First note that no matter what is the diffusion coefficient, for an infinitesimal density gradient, to leading order the density profile is linear:
\begin{align}
\rho\left(\vec{r},t\right)=\rho_{L}+\frac{r_{x}\epsilon}{L_{x}}+o\left(\frac{\epsilon}{L_{x}}\right) .\label{linearap}
\end{align}
For any group of sites $G$ we may write an evolution equation
\begin{align}
&\frac{\partial}{\partial t}\left\langle\prod_{\vec{r}\in G}n_{\alpha}\left(\vec{r},t\right)\right\rangle=\nonumber\\
&\left\langle\sum_{\vec{r}\in G}\prod_{\begin{array}{c}\vec{r}'\in G\\\vec{r}'\neq\vec{r}\end{array}}n_{\alpha}\left(\vec{r}',t\right)\sum_{\begin{array}{c}\hat{d}\\\vec{r}+\hat{d}\notin G\end{array}}\right.\nonumber\\
&\left.\left[n_{\alpha}\left(\vec{r}+\hat{d},t\right)-n_{\alpha}\left(\vec{r},t\right)\right]K_{\alpha,\hat{d}}\left(\vec{r},t\right)\right\rangle .\label{evolutiongen}
\end{align}
We now assume that for any group $G$ the correlations between the relevant sites, $C_{G}$, are at most of order $\epsilon/L_{x}$
\begin{align}
C_{G}=\frac{\epsilon}{L_{x}}C^{(1)}_{G}+o\left(\frac{\epsilon}{L_{x}}\right) .
\end{align}
Setting this assumption in Eq. (\ref{evolutiongen}) in the steady state for the groups which comprise $K_{\alpha,\hat{x}}\left(\vec{r},t\right)$ in the $m=2$ KA model (for example) yields
\begin{align}
&0=\left\{\rho^{4}\left(\vec{r},t\right)\left[1-\rho\left(\vec{r},t\right)\right]\left[1-\rho^{3}\left(\vec{r},t\right)\right]+C^{(1)}_{K}\right\}\frac{\epsilon}{L_{x}}+\nonumber\\
&+o\left(\frac{\epsilon}{L_{x}}\right) ,
\end{align}
where $C^{(1)}_{K}$ is the sum of all the first order terms in the correlators of the groups comrpising $K_{\alpha,\hat{x}}\left(\vec{r},t\right)$. For general values of $\rho\left(\vec{r},t\right)$ we find that the correlator is not zero, $C^{(1)}_{K}\neq0$. Hence, the correlations between some groups of sites is of order $\epsilon/L_{x}$. Note that this does not contradict~\cite{129Aminov2015}, where it was shown that correlations between two sites are of order $\epsilon^{2}/L_{x}$, since here we consider correlations between more than two sites. As a further note, we believe that the correlators can be analytically calculated using the formalism introduced in~\cite{Derrida1993}, but this is beyond the scope of this paper.

For gradient models, we want to show that $C^{(1)}_{\vec{r},\hat{d}}=0$ (see Eq. (\ref{corrapp})). In order to do this, we note that by definition the current in gradient models may be written as
\begin{align}
J=f\left(\vec{r}+\hat{d}\right)-f\left(\vec{r}\right) ,\label{currentgrad}
\end{align}
where $f\left(\vec{r}\right)$ is some model-dependent function of the occupancy of sites in the neighborhood of the site $\vec{r}$. In the limit $\frac{\epsilon}{L_{x}}\ll1$ we may expand $f\left(\vec{r}\right)$ by
\begin{align}
f\left(\vec{r}\right)=\bar{f}\left(\vec{r}\right)+\frac{\epsilon}{L_{x}}C_{f}\left(\vec{r}\right)+o\left(\frac{\epsilon}{L_{x}}\right) ,
\end{align}
where $\bar{f}\left(\vec{r}\right)$ is the equilibrium value of $f\left(\vec{r}\right)$. Therefore, the current, Eq. (\ref{currentgrad}), may be written as
\begin{align}
&J=\bar{f}\left(\vec{r}+\hat{d}\right)-\bar{f}\left(\vec{r}\right)+\nonumber\\
&+\frac{\epsilon}{L_{x}}\left[C_{f}\left(\vec{r}+\hat{d}\right)-C_{f}\left(\vec{r}\right)\right]+o\left(\frac{\epsilon}{L_{x}}\right) .
\end{align}
We therefore identify in gradient models that
\begin{align}
C^{(1)}_{\vec{r},\hat{x}}=\lim_{\frac{\epsilon}{L_{x}}\rightarrow0}C_{f}\left(\vec{r}+\hat{d}\right)-C_{f}\left(\vec{r}\right) .
\end{align}
In the joint limit $\epsilon\ll1$ and $L\gg1$, the density gradients are of order $\frac{\epsilon}{L_{x}}$ and thus the gradient of any function that depends on the density, and does not depend explicitly on $\frac{\epsilon}{L_{x}}$ is also of order $\frac{\epsilon}{L_{x}}$. Therefore
\begin{align}
C_{f}\left(\vec{r}+\hat{d}\right)-C_{f}\left(\vec{r}\right)=O\left(\frac{\epsilon}{L_{x}}\right)
\end{align}
and we conclude that $C^{(1)}_{\vec{r},\hat{x}}=0$. This means that in gradient models, the correlations that appear when the system is driven out of equilibrium, even if they are of order $O\left(\frac{\epsilon}{L_{x}}\right)$, do not affect the first order term of the current, since their main contribution is via their discrete derivative which is of order $O\left(\frac{\epsilon^{2}}{L^{2}_{x}}\right)$. Thus, the first order term of the current is determined solely by the behavior of the system at equilibrium, $\bar{f}$. Hence, the no-correlation derivation is an approximation for non-gradient models, and an exact derivation for gradient models.

\section{Fluctuations}
\label{sec_fluct}

Until now, we discussed how the diffusion coefficient is related to the average current in the system when it is driven out of equilibrium. In fact, from the fluctuation-dissipation theorem, one finds that the diffusion is also related to the fluctuations in the current when the system is in equilibrium~\cite{Krapivsky2012}
\begin{align}
D\left(\rho\right)=\frac{1}{2}\frac{d^{2}F}{d\rho^{2}}\lim_{\begin{array}{c}t\rightarrow\infty\\L_{x}\rightarrow\infty\end{array}}\frac{L_{x}\left\langle J^{2}_{\alpha}(t)\right\rangle}{t} ,
\end{align}
where $F$ is the free energy of the system, which for non-interacting lattice gases (such as KCMs), is given by~\cite{Krapivsky2012}
\begin{align}
F=\rho\ln\rho+\left(1-\rho\right)\ln\left(1-\rho\right) ,
\end{align}
and $J_{\alpha}(t)$ is the integrated current in the system from time $0$ to time $t$ under trajectory $\alpha$ when the system is in equilibrium. Thus, for KCMs the diffusion coefficient may be written as
\begin{align}
D\left(\rho\right)=\frac{1}{2\rho\left(1-\rho\right)}\lim_{\begin{array}{c}t\rightarrow\infty\\L_{x}\rightarrow\infty\end{array}}\frac{L_{x}\left\langle J^{2}_{\alpha}(t)\right\rangle}{t} .
\end{align}

\begin{figure}
\begin{centering}
\includegraphics[width=0.8\columnwidth]{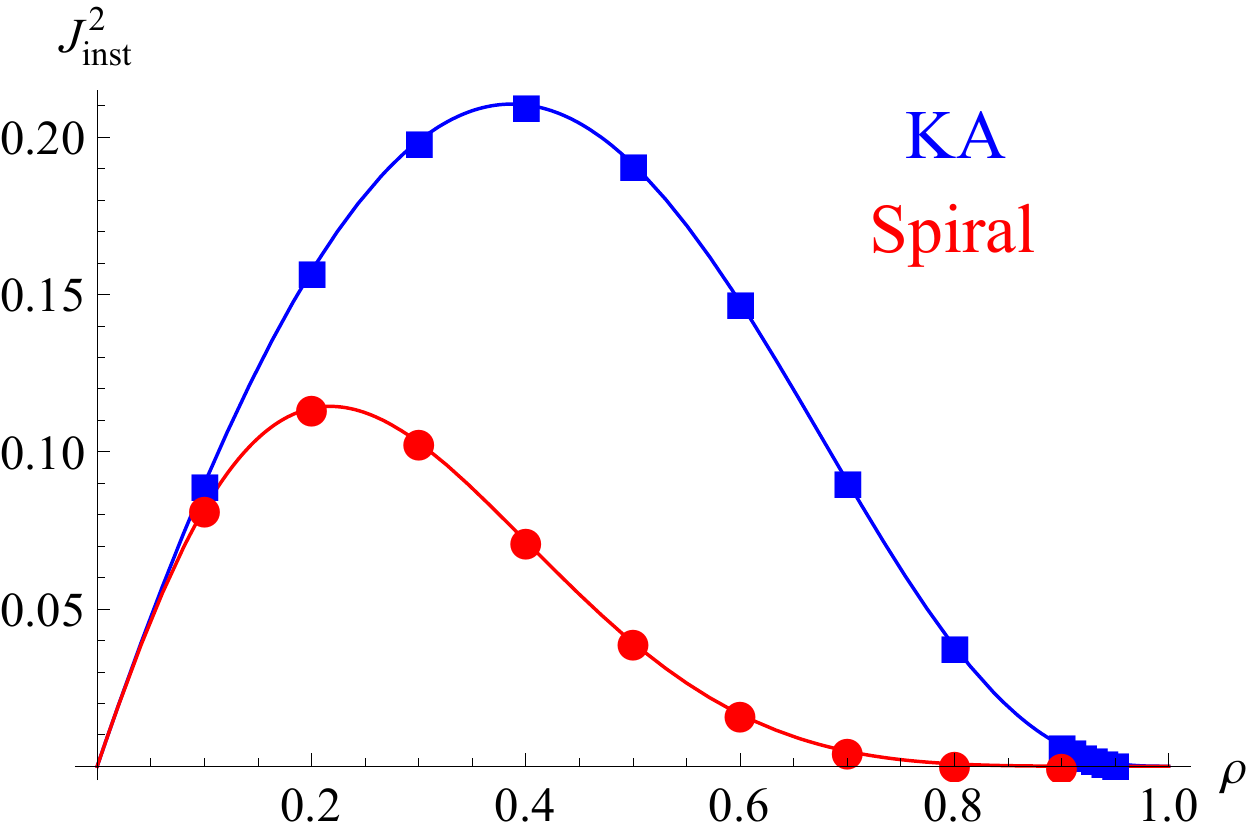}
\par\end{centering}
\caption{The fluctuations in the instantaneous current $\left\langle j^{2}_{inst}\right\rangle$ vs. the density $\rho$. The symbols are simulation results and the continuous lines are the analytical results, Eq. (\ref{fluct}).}
\label{fluctfig}
\end{figure}

Measuring the fluctuations in the current in the proper limits is computationally demanding. Even though by construction there are no correlations between the occupancies of different sites at the same time in KCMs, there are long time dynamical heterogeneities which means that there are correlations between the occupancy of different sites at different times, and therefore also of the current at different times. Instead of measuring the long time fluctuations, we measure a related quantity, the fluctuations in the instantaneous current at equilibrium $\left\langle j^{2}_{inst}\right\rangle$. Namely, at each time step of the simulation we measure the instantaneous current, and then average over the fluctuations. At each time step, either a move occurs and thus $j^{2}_{inst}=1$ or a move does not occur and then $j^{2}_{inst}=0$. Since the system is in equilibrium, there is no correlation between the occupancy of different sites and thus $\left\langle j^{2}_{inst}\right\rangle$ is equal to the probability that a move occurs
\begin{align}
\left\langle j^{2}_{inst}\right\rangle=\rho\left(1-\rho\right)D_{NC}\left(\rho\right) .\label{fluct}
\end{align}
Figure \ref{fluctfig} shows the excellent agreement between the numerical measurement of $\left\langle j^{2}_{inst}\right\rangle$ and Eq. (\ref{fluct}).

The main conclusion here is that although in equilibrium there are no correlations between the occupancy of different sites \textit{at the same time}, there are correlations between the occupancy of different sites \textit{at different time}. These temporal correlations are related to the spatial, same time, non-negligible correlations that appear when the system is driven out of equilibrium. The origin of these correlations in equilibrium is the dynamic heterogeneities, one of the main characteristics of KCMs in general.

\section{Finding the diffusion coefficient numerically}
\label{sec_num}

Now that we know that the diffusion coefficient is different from the result of the NC approximation, we want to find it numerically. To do this we performed two types of simulations; one involves a small difference in the densities of the two reservoirs, and the second involves a large difference in the densities. In the first method, we use the definition of the diffusion coefficient, Eq. (\ref{diffdef}), directly and simulate a system connected to two reservoirs with an infinitesimal density difference $\epsilon\ll1$, and measure the current in the system. We than decrease $\epsilon$ and increase the system size $L_{x}$ until the results converge. Theoretically, the two limits commute, but in practice if $\epsilon$ is too small, the density gradient will be smaller than the finite size effects and the statistical fluctuations. In the second type of simulations, as described in Section \ref{secdenprof} above we consider a system connected to two reservoirs, one with density $\rho_{L}=0$ and the other with a finite density $\rho_{R}=\rho_{0}$. We let the system evolve and measure the steady state density profile $\rho(x)$. In order to get the diffusion coefficient from the steady state density profile, we integrate Eq. (\ref{diffeq}) in the steady state over $x$
\begin{align}
0=D\left(\rho(x)\right)\frac{d\rho(x)}{dx}-D\left(\rho(0)\right)\frac{d\rho(0)}{dx} .
\end{align}
Since $\rho(0)=0$, we can arbitrarily set $D(0)=1$, and thus
\begin{align}
D\left(\rho\right)=\frac{\frac{d\rho(0)}{dx}}{\frac{d\rho(x)}{dx}} .\label{dfromder}
\end{align}
Note that Eq. (\ref{dfromder}) does not depend on $\rho_{0}$.

\begin{figure}
\begin{centering}
\includegraphics[width=0.8\columnwidth]{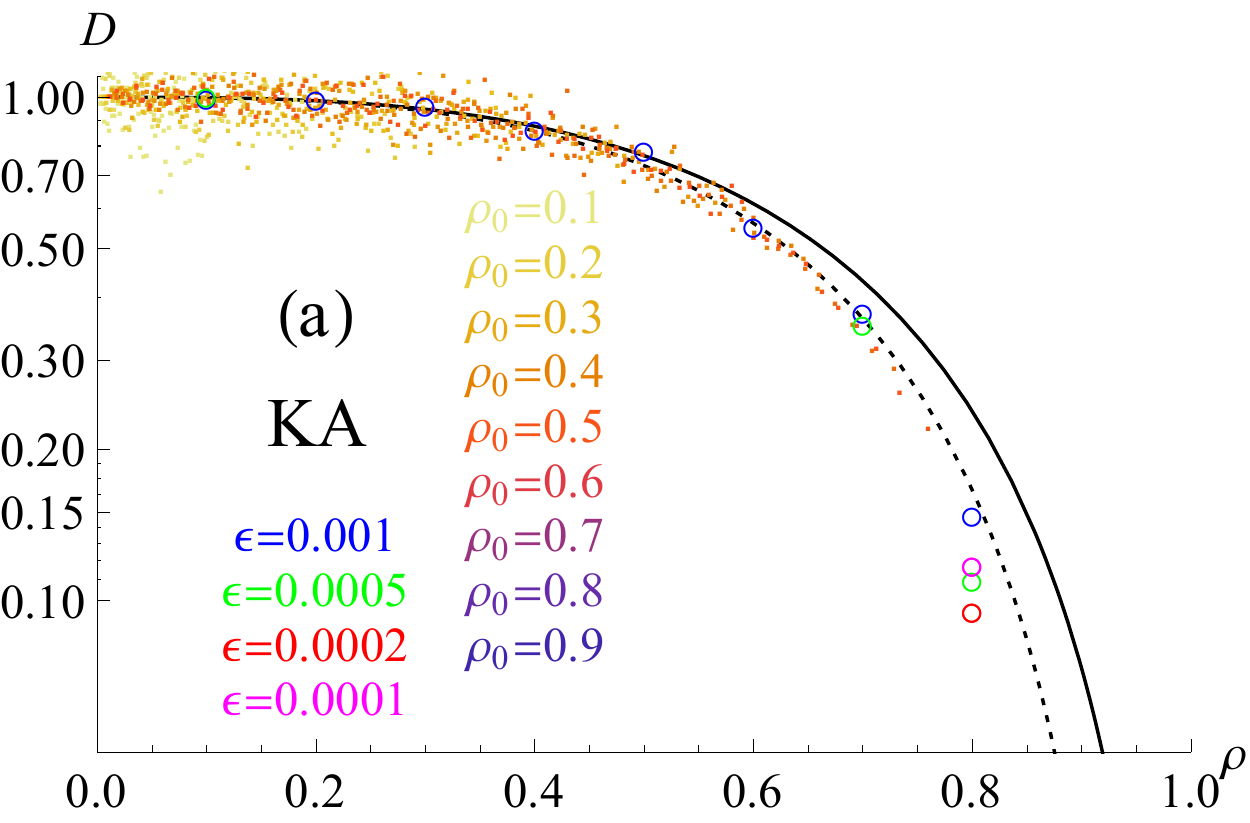}\\
\includegraphics[width=0.8\columnwidth]{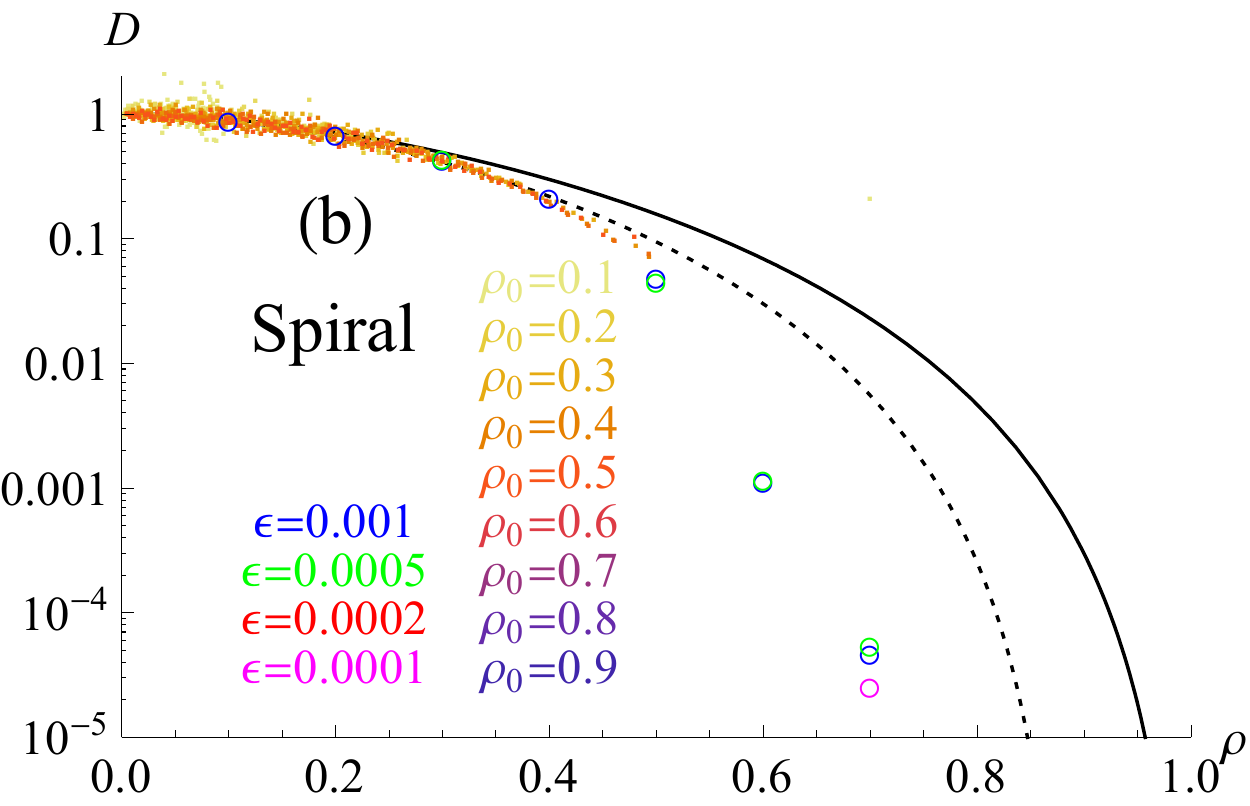}
\par\end{centering}
\caption{A comparison between the diffusion coefficient derived by the approximation $D_{NC}$ (black continuous line), Eqs. (\ref{diffka2}) and (\ref{diffsp}), by the approximation $D_{NC+\rho_{eff}}$ (black dotted line), Eq. (\ref{deff}), by the derivative of the density profile for various $\rho_{0}$ (small full squares), Eq. (\ref{dfromder}), and by the small gradient for various $\epsilon$ (larger empty circles), Eq. (\ref{diffdef}).}
\label{difffin}
\end{figure}

Figure \ref{difffin} shows the diffusion coefficient derived from these two methods, and the approximated diffusion coefficient, Eqs. (\ref{diffka2}) for the KA model and (\ref{diffsp}) for the spiral model. We first note that $D$ as derived from Eq. (\ref{dfromder}) indeed does not depend on $\rho_{0}$, and that it agrees with the first method of deriving the diffusion coefficient. This supports the fact that the deviations we observe from the NC result are physical and are not sensitive to the numerical method. Second, we note that as expected, the diffusion coefficient is smaller than the NC approximation.

\subsection*{Effective reservoir density approximation}

\begin{figure}
\begin{centering}
\includegraphics[width=0.8\columnwidth]{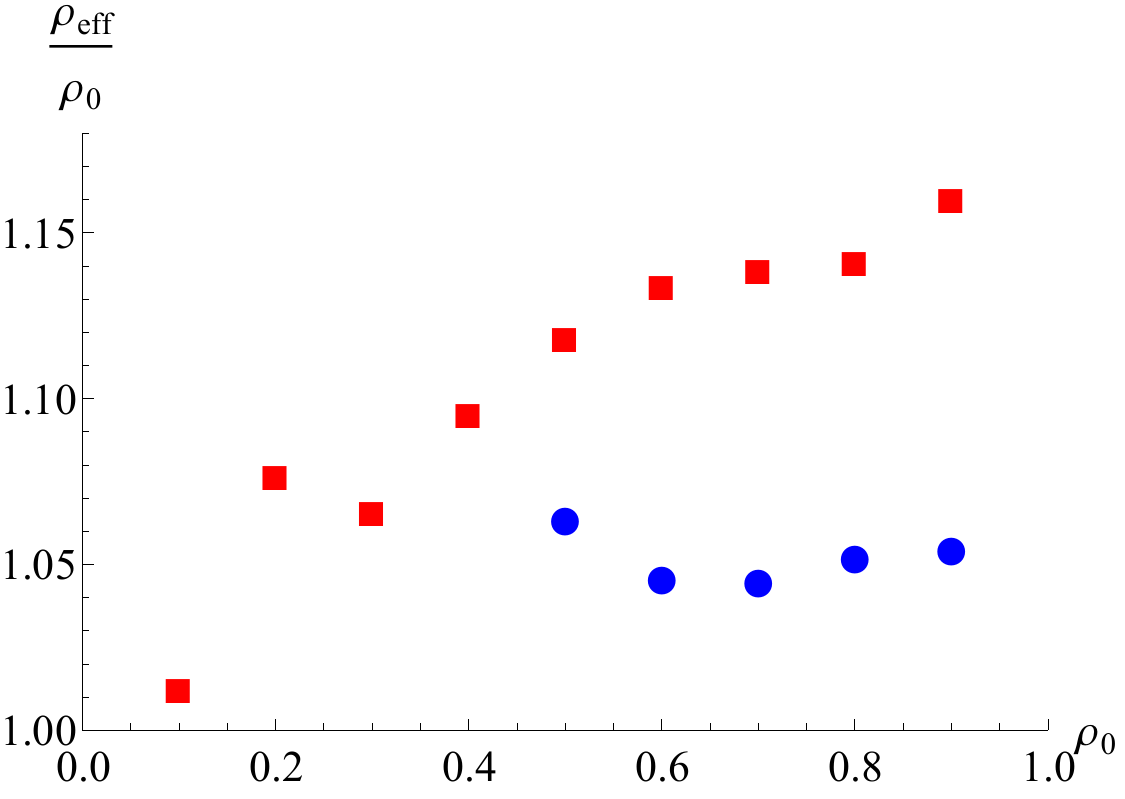}
\par\end{centering}
\caption{The normalized effective density vs. $\rho_{0}$ for the spiral model (red squares) and the KA model (blue circles). No data is shown for the KA model at $\rho_{0}\leq0.4$ since there it is very hard to resolve the difference.}
\label{scaling}
\end{figure}

A peculiar phenomenon that we observed numerically is that for each value of the reservoir density $\rho_0$ there is some effective density $\rho_{eff}$ such that the density profile in the steady state may be very well approximated by
\begin{align}
\rho\left(x;\rho_{0}\right)=\frac{\rho_{0}}{\rho_{eff}}\rho_{NC}\left(x;\rho_{eff}\right) ,\label{rhoscale}
\end{align}
where $\rho_{NC}\left(x;\rho_{eff}\right)$ is the steady state solution of the diffusion equation with $D=D_{NC}$ and the boundary conditions $\rho_{L}=0$ and $\rho_{R}=\rho_{eff}$. This is shown in Fig. \ref{mdif1}a for the KA model and in Fig. \ref{mdif1}b for the spiral model for several densities, but it works in all the densities we checked in both the KA and the spiral models. Note that $\rho_{eff}$ can be higher than $1$. Figure \ref{scaling} shows the value of $\rho_{eff}$ vs. $\rho_{0}$. Even in the spiral model, where the difference is more pronounced, we find that the relative difference is smaller than $0.2$. Based on this observation, we now find an approximation for the diffusion coefficient, $D_{NC+\rho_{eff}}$.

Since $\rho\left(x;\rho_{0}\right)$ is monotonic with respect to $x$ we can invert it, and thus Eq. (\ref{rhoscale}) is equivalent to
\begin{align}
x\left(z\rho_{0};\rho_{0}\right)=x\left(z\rho_{eff};\rho_{eff}\right) ,\label{xscaling}
\end{align}
where $z\in\left[0,1\right]$. Combining Eqs. (\ref{2int}) and (\ref{c1def}) yields
\begin{align}
\frac{\int^{z\rho_{0}}_{0}D\left(\rho\right)d\rho}{\int^{\rho_{0}}_{0}D\left(\rho\right)d\rho}=\frac{\int^{z\rho_{eff}}_{0}D_{NC}\left(\rho\right)d\rho}{\int^{\rho_{eff}}_{0}D_{NC}\left(\rho\right)d\rho} ,\label{zdif}
\end{align}
because the left hand side of Eq. (\ref{xscaling}) is found by the true diffusion coefficient $D$, and the right hand side is found by the approximate diffusion coefficient $D_{NC}$. We now change the integration variable in the integrals on the right hand side to
\begin{align}
\rho'=\rho\frac{\rho_{0}}{\rho_{eff}} ,
\end{align}
such that
\begin{align}
\frac{\int^{z\rho_{0}}_{0}D\left(\rho\right)d\rho}{\int^{\rho_{0}}_{0}D\left(\rho\right)d\rho}=\frac{\int^{z\rho_{0}}_{0}D_{NC}\left(\rho'\frac{\rho_{eff}}{\rho_{0}}\right)d\rho'}{\int^{\rho_{0}}_{0}D_{NC}\left(\rho'\frac{\rho_{eff}}{\rho_{0}}\right)d\rho'} .
\end{align}
Since the limits on the integrals on both sides are the same, and the functions are monotonic, we may deduce that the integrands are the same
\begin{align}
D\left(\rho\right)=D_{NC}\left(\rho\frac{\rho_{eff}}{\rho_{0}}\right)\equiv D_{NC+\rho_{eff}}\left(\rho\right) .\label{deff}
\end{align}
Now, since $D\left(\rho\right)$ does not depend on $\rho_{0}$ we deduce that $\rho_{eff}$ is a linear function of $\rho_{0}$, with a model-dependent slope $s$. Based on Fig. \ref{scaling}, and considering only the higher values of $\rho_{0}$ where the difference is more pronounced, we estimate this slope as $s_{KA}=1.05$  and $s_{SP}=1.13$. However, this means that $D(\rho=1/s)=D_{NC}(1)=0$, which we do not see happening in the simulations ($1/s_{KA}\approx0.95$ and $1/s_{SP}\approx0.88$, and we considered $\rho_{0}$ up to $0.9$). For the KA model, the value of $0.95$ is suspiciously close to the critical density of square systems of size $L=100$ (as in the simulations), but since this phenomenon occurs even at rather low densities we do not expect any true relation. Thus, we conclude that the seemingly equality in Eq. (\ref{rhoscale}) is actually just a very good approximation, and therefore Eq. (\ref{deff}) is also just a very good approximation.

\section{Summary}
\label{sec_summary}

In this paper we showed that lattice-gas kinetically-constrained models can be coarse grained to a hydrodynamic description with a non-trivial and non-linear diffusion coefficient. This diffusion coefficient is measured by driving the system out of equilibrium by an infinitesimally-small density gradient. We showed that in general KCMs, this diffusion coefficient is well approximated by a polynomial of the density which can be easily obtained from the kinetic constraints. However, the exact value of the diffusion coefficient depends also on infinitesimally small spatial correlations between sites that appear as the system is driven out of equilibrium due to even infinitesimally small density gradients. From the fluctuation-dissipation theorem, we may infer then that the fluctuations (derived from long-time measurements at equilibrium), which are related to the dissipation (derived from instantaneous measurements out of equilibrium), are also non-trivial. Thus, we may use the measurements of the diffusion coefficient to evaluate very long time correlations in equilibrium, which are non-trivial and not easy to measure since the relaxation time in KCMs is extremely long.

These correlations are strongly related to the correlations that appear as the system is driven out of equilibrium by an external field \cite{Sellitto2008,130Turci2012a}. They are not exactly the same, since here the control parameter is the particle density, while in \cite{Sellitto2008,130Turci2012a} the control parameters are both the density and the strength of the applied field. Although they are different, they both arise due to the breaking of inversion symmetry, and it would be interesting to study possible general relations between the diffusion coefficient and the field-induced current.

Finding an exact analytical expression for the diffusion coefficient is a promising research direction which would shed light on the dynamics of these models. Two possible approaches are to investigate very small systems and exactly solve the master equation for them, or to use the formalism introduced in \cite{Derrida1993}, in which the correlations are expressed as a multiplication of model-specific left and right vectors, representing the two reservoirs, and an infinite number of matrices, representing the local transition matrices for single sites. In both approaches, the calculations would be less cumbersome if a relatively simple model is investigated; We propose a variation of the model used in \cite{127Goncalves2009}, in which a particle can move on a one-dimensional lattice from site $r$ to site $r+1$ if site $r+1$ is vacant, and if either site $r-1$ is vacant or site $r+2$ is vacant. The difference between this proposed model and the one used in \cite{127Goncalves2009} is that the hopping rate is either $0$ if both sites $r-1$ and $r+2$ are occupied, and $1$ otherwise. Unlike the model used in \cite{127Goncalves2009}, our proposed model is non-gradient, and we thus expect the diffusion coefficient to differ from the NC approximation.

In the two-dimensional models we investigated here numerically, the $m=2$ KA model and the spiral model, the diffusion coefficient is continuous with respect to the density. We suspect, however, that using the same setup in our extension of the spiral model to three dimensions~\cite{085Ghosh2014} will create some interesting phenomena. In that model, when the density is above the critical density of two-dimensional directed percolation, there are frozen quasi-$1D$ clusters in each of the $y-z$ planes perpendicular to the density gradient, which we assumed is in the $x$ direction, even for periodic boundary conditions. Several questions come to mind: do these clusters cause the system to freeze, or are other particles able to travel between these frozen clusters, similarly to what happens in this model when it is not connected to reservoirs~\cite{086Segall2016}? If the particles can travel between the frozen clusters, is there some higher density at which the diffusion stops? Is there a singularity in the diffusion coefficient or its derivative with respect to the density? Is there some density at which the motion becomes sub-diffusive? When the initial condition is a completely empty system, these clusters cannot appear due to the microscopic reversibility of the dynamics, so what happens then? These questions can guide future research on this topic.

\section*{Acknowledgments}

We thank Cristina Toninelli, Oriane Blondel, Juan Garrahan, Robert Jack, Fabio Leoni, Nimrod Segall, and Hadas Shem-Tov for helpful discussions. We especially thank Baruch Meerson and Paul Krapivsky for critically reviewing this research. This research was supported by the Israel Science Foundation grant No. 968/12 and by the Prof A Pazy Research Foundation.

\bibliography{references}

\begin{thebibliography}{48}
\expandafter\ifx\csname natexlab\endcsname\relax\def\natexlab#1{#1}\fi
\expandafter\ifx\csname bibnamefont\endcsname\relax
  \def\bibnamefont#1{#1}\fi
\expandafter\ifx\csname bibfnamefont\endcsname\relax
  \def\bibfnamefont#1{#1}\fi
\expandafter\ifx\csname citenamefont\endcsname\relax
  \def\citenamefont#1{#1}\fi
\expandafter\ifx\csname url\endcsname\relax
  \def\url#1{\texttt{#1}}\fi
\expandafter\ifx\csname urlprefix\endcsname\relax\def\urlprefix{URL }\fi
\providecommand{\bibinfo}[2]{#2}
\providecommand{\eprint}[2][]{\url{#2}}

\bibitem[{\citenamefont{Ritort and Sollich}(2003)}]{017Ritort2003}
\bibinfo{author}{\bibfnamefont{F.}~\bibnamefont{Ritort}} \bibnamefont{and}
  \bibinfo{author}{\bibfnamefont{P.}~\bibnamefont{Sollich}},
  \bibinfo{journal}{Advances in Physics} \textbf{\bibinfo{volume}{52}},
  \bibinfo{pages}{219} (\bibinfo{year}{2003}),
  \urlprefix\url{http://dx.doi.org/10.1080/0001873031000093582}.

\bibitem[{\citenamefont{Garrahan et~al.}(2010)\citenamefont{Garrahan, Sollich,
  and Toninelli}}]{018Garrahan2010}
\bibinfo{author}{\bibfnamefont{J.}~\bibnamefont{Garrahan}},
  \bibinfo{author}{\bibfnamefont{P.}~\bibnamefont{Sollich}}, \bibnamefont{and}
  \bibinfo{author}{\bibfnamefont{C.}~\bibnamefont{Toninelli}},
  \emph{\bibinfo{title}{{Dynamical Heterogeneities in Glasses, Colloids, and
  Granular Media}}}, International Series of Monographs on Physics
  (\bibinfo{publisher}{Oxford University Press}, \bibinfo{year}{2010}),
  \urlprefix\url{http://arxiv.org/abs/1009.6113}.

\bibitem[{\citenamefont{Berthier and
  Garrahan}(2003{\natexlab{a}})}]{100Berthier2003a}
\bibinfo{author}{\bibfnamefont{L.}~\bibnamefont{Berthier}} \bibnamefont{and}
  \bibinfo{author}{\bibfnamefont{J.}~\bibnamefont{Garrahan}},
  \bibinfo{journal}{Physical Review E} \textbf{\bibinfo{volume}{68}},
  \bibinfo{pages}{041201} (\bibinfo{year}{2003}{\natexlab{a}}),
  \urlprefix\url{http:/dx.doi.org/10.1103/PhysRevE.68.041201}.

\bibitem[{\citenamefont{Chandler et~al.}(2006)\citenamefont{Chandler, Garrahan,
  Jack, Maibaum, and Pan}}]{Chandler2006}
\bibinfo{author}{\bibfnamefont{D.}~\bibnamefont{Chandler}},
  \bibinfo{author}{\bibfnamefont{J.}~\bibnamefont{Garrahan}},
  \bibinfo{author}{\bibfnamefont{R.}~\bibnamefont{Jack}},
  \bibinfo{author}{\bibfnamefont{L.}~\bibnamefont{Maibaum}}, \bibnamefont{and}
  \bibinfo{author}{\bibfnamefont{A.}~\bibnamefont{Pan}},
  \bibinfo{journal}{Physical Review E} \textbf{\bibinfo{volume}{74}},
  \bibinfo{pages}{051501} (\bibinfo{year}{2006}),
  \urlprefix\url{http://dx.doi.org/10.1103/PhysRevE.74.051501}.

\bibitem[{\citenamefont{L{\'{e}}onard and Harrowell}(2010)}]{055Leonard2010}
\bibinfo{author}{\bibfnamefont{S.}~\bibnamefont{L{\'{e}}onard}}
  \bibnamefont{and}
  \bibinfo{author}{\bibfnamefont{P.}~\bibnamefont{Harrowell}},
  \bibinfo{journal}{Journal of Chemical Physics}
  \textbf{\bibinfo{volume}{133}}, \bibinfo{pages}{244502}
  (\bibinfo{year}{2010}), \urlprefix\url{http://dx.doi.org/10.1063/1.3511721}.

\bibitem[{\citenamefont{Pastore et~al.}(2013)\citenamefont{Pastore, Ciamarra,
  and Coniglio}}]{106Pastore2013}
\bibinfo{author}{\bibfnamefont{R.}~\bibnamefont{Pastore}},
  \bibinfo{author}{\bibfnamefont{M.}~\bibnamefont{Ciamarra}}, \bibnamefont{and}
  \bibinfo{author}{\bibfnamefont{A.}~\bibnamefont{Coniglio}},
  \bibinfo{journal}{Fractals} \textbf{\bibinfo{volume}{21}},
  \bibinfo{pages}{1350021} (\bibinfo{year}{2013}),
  \urlprefix\url{http://dx.doi.org/10.1142/S0218348X13500217}.

\bibitem[{\citenamefont{Pastore et~al.}(2015)\citenamefont{Pastore, Coniglio,
  and Cimarra}}]{Pastore2015a}
\bibinfo{author}{\bibfnamefont{R.}~\bibnamefont{Pastore}},
  \bibinfo{author}{\bibfnamefont{A.}~\bibnamefont{Coniglio}}, \bibnamefont{and}
  \bibinfo{author}{\bibfnamefont{M.}~\bibnamefont{Cimarra}},
  \bibinfo{journal}{Scientific Reports} \textbf{\bibinfo{volume}{5}},
  \bibinfo{pages}{11770} (\bibinfo{year}{2015}),
  \urlprefix\url{http://dx.doi.org/10.1038/srep11770}.

\bibitem[{\citenamefont{Garrahan and Chandler}(2002)}]{Garrahan2002a}
\bibinfo{author}{\bibfnamefont{J.~P.} \bibnamefont{Garrahan}} \bibnamefont{and}
  \bibinfo{author}{\bibfnamefont{D.}~\bibnamefont{Chandler}},
  \bibinfo{journal}{Physical Review Letters} \textbf{\bibinfo{volume}{89}},
  \bibinfo{pages}{035704} (\bibinfo{year}{2002}),
  \urlprefix\url{http://link.aps.org/doi/10.1103/PhysRevLett.89.035704}.

\bibitem[{\citenamefont{Berthier and
  Garrahan}(2003{\natexlab{b}})}]{101Berthier2003}
\bibinfo{author}{\bibfnamefont{L.}~\bibnamefont{Berthier}} \bibnamefont{and}
  \bibinfo{author}{\bibfnamefont{J.}~\bibnamefont{Garrahan}},
  \bibinfo{journal}{Journal of Chemical Physics}
  \textbf{\bibinfo{volume}{119}}, \bibinfo{pages}{4367}
  (\bibinfo{year}{2003}{\natexlab{b}}),
  \urlprefix\url{http://dx.doi.org/10.1063/1.1593020}.

\bibitem[{\citenamefont{Marinari and Pitard}(2005)}]{104Marinari2005}
\bibinfo{author}{\bibfnamefont{E.}~\bibnamefont{Marinari}} \bibnamefont{and}
  \bibinfo{author}{\bibfnamefont{E.}~\bibnamefont{Pitard}},
  \bibinfo{journal}{Europhysics Letters} \textbf{\bibinfo{volume}{69}},
  \bibinfo{pages}{235} (\bibinfo{year}{2005}),
  \urlprefix\url{http://dx.doi.org/10.1209/epl/i2004-10323-2}.

\bibitem[{\citenamefont{Whitelam et~al.}(2005)\citenamefont{Whitelam, Berthier,
  and Garrahan}}]{Whitelam2005}
\bibinfo{author}{\bibfnamefont{S.}~\bibnamefont{Whitelam}},
  \bibinfo{author}{\bibfnamefont{L.}~\bibnamefont{Berthier}}, \bibnamefont{and}
  \bibinfo{author}{\bibfnamefont{J.}~\bibnamefont{Garrahan}},
  \bibinfo{journal}{Physical Review E} \textbf{\bibinfo{volume}{71}},
  \bibinfo{pages}{026128} (\bibinfo{year}{2005}),
  \urlprefix\url{http://link.aps.org/doi/10.1103/PhysRevE.71.026128}.

\bibitem[{\citenamefont{Berthier et~al.}(2007)\citenamefont{Berthier, Biroli,
  Bouchaud, Kob, Miyazaki, and Reichman}}]{105Berthier2007}
\bibinfo{author}{\bibfnamefont{L.}~\bibnamefont{Berthier}},
  \bibinfo{author}{\bibfnamefont{G.}~\bibnamefont{Biroli}},
  \bibinfo{author}{\bibfnamefont{J.}~\bibnamefont{Bouchaud}},
  \bibinfo{author}{\bibfnamefont{W.}~\bibnamefont{Kob}},
  \bibinfo{author}{\bibfnamefont{K.}~\bibnamefont{Miyazaki}}, \bibnamefont{and}
  \bibinfo{author}{\bibfnamefont{D.}~\bibnamefont{Reichman}},
  \bibinfo{journal}{Journal of Chemical Physics}
  \textbf{\bibinfo{volume}{126}}, \bibinfo{pages}{184504}
  (\bibinfo{year}{2007}), \urlprefix\url{http://dx.doi.org/10.1063/1.2721555}.

\bibitem[{\citenamefont{Teomy and Shokef}(2015)}]{087Teomy2015}
\bibinfo{author}{\bibfnamefont{E.}~\bibnamefont{Teomy}} \bibnamefont{and}
  \bibinfo{author}{\bibfnamefont{Y.}~\bibnamefont{Shokef}},
  \bibinfo{journal}{Physical Review E} \textbf{\bibinfo{volume}{92}},
  \bibinfo{pages}{032133} (\bibinfo{year}{2015}),
  \urlprefix\url{http://dx.doi.org/10.1103/PhysRevE.92.032133}.

\bibitem[{\citenamefont{Pastore et~al.}(2016)\citenamefont{Pastore, de~Candia,
  Fierro, Pica~Ciamarra, and Coniglio}}]{Pastore2016}
\bibinfo{author}{\bibfnamefont{R.}~\bibnamefont{Pastore}},
  \bibinfo{author}{\bibfnamefont{A.}~\bibnamefont{de~Candia}},
  \bibinfo{author}{\bibfnamefont{A.}~\bibnamefont{Fierro}},
  \bibinfo{author}{\bibfnamefont{M.}~\bibnamefont{Pica~Ciamarra}},
  \bibnamefont{and} \bibinfo{author}{\bibfnamefont{A.}~\bibnamefont{Coniglio}},
  \bibinfo{journal}{Journal of Statistical Mechanics: Theory and Experiments}
  \textbf{\bibinfo{volume}{2016}}, \bibinfo{pages}{74011}
  (\bibinfo{year}{2016}),
  \urlprefix\url{http://dx.doi.org/10.1088/1742-5468/2016/7/074011}.

\bibitem[{\citenamefont{Butler and Harrowell}(1991)}]{114Butler1991}
\bibinfo{author}{\bibfnamefont{S.}~\bibnamefont{Butler}} \bibnamefont{and}
  \bibinfo{author}{\bibfnamefont{P.}~\bibnamefont{Harrowell}},
  \bibinfo{journal}{Journal of Chemical Physics} \textbf{\bibinfo{volume}{95}},
  \bibinfo{pages}{4454} (\bibinfo{year}{1991}),
  \urlprefix\url{http://dx.doi.org/10.1063/1.461768}.

\bibitem[{\citenamefont{Kob and Andersen}(1993)}]{067Kob1993}
\bibinfo{author}{\bibfnamefont{W.}~\bibnamefont{Kob}} \bibnamefont{and}
  \bibinfo{author}{\bibfnamefont{H.}~\bibnamefont{Andersen}},
  \bibinfo{journal}{Physical Review E} \textbf{\bibinfo{volume}{48}},
  \bibinfo{pages}{4364} (\bibinfo{year}{1993}),
  \urlprefix\url{http://dx.doi.org/10.1103/PhysRevE.48.4364}.

\bibitem[{\citenamefont{Einax and Schulz}(2001)}]{094Einax2001}
\bibinfo{author}{\bibfnamefont{M.}~\bibnamefont{Einax}} \bibnamefont{and}
  \bibinfo{author}{\bibfnamefont{M.}~\bibnamefont{Schulz}},
  \bibinfo{journal}{Journal of Chemical Physics}
  \textbf{\bibinfo{volume}{115}}, \bibinfo{pages}{2282} (\bibinfo{year}{2001}),
  \urlprefix\url{http://dx.doi.org/10.1063/1.1383053}.

\bibitem[{\citenamefont{Schulz and Trimper}(2002)}]{099Schulz2002}
\bibinfo{author}{\bibfnamefont{M.}~\bibnamefont{Schulz}} \bibnamefont{and}
  \bibinfo{author}{\bibfnamefont{S.}~\bibnamefont{Trimper}},
  \bibinfo{journal}{Journal of Physics: Condensed Matter}
  \textbf{\bibinfo{volume}{14}}, \bibinfo{pages}{1437} (\bibinfo{year}{2002}),
  \urlprefix\url{http://iopscience.iop.org/0953-8984/14/7/304}.

\bibitem[{\citenamefont{Kuhlmann et~al.}(2005)\citenamefont{Kuhlmann, Trimper,
  and Schulz}}]{095Kuhlmann2005}
\bibinfo{author}{\bibfnamefont{C.}~\bibnamefont{Kuhlmann}},
  \bibinfo{author}{\bibfnamefont{S.}~\bibnamefont{Trimper}}, \bibnamefont{and}
  \bibinfo{author}{\bibfnamefont{M.}~\bibnamefont{Schulz}},
  \bibinfo{journal}{Physica Status Solidi (B)} \textbf{\bibinfo{volume}{242}},
  \bibinfo{pages}{2401} (\bibinfo{year}{2005}),
  \urlprefix\url{http://dx.doi.org/10.1002/pssb.200440092}.

\bibitem[{\citenamefont{Jung et~al.}(2005)\citenamefont{Jung, Garrahan, and
  Chandler}}]{102Jung2005}
\bibinfo{author}{\bibfnamefont{Y.}~\bibnamefont{Jung}},
  \bibinfo{author}{\bibfnamefont{J.}~\bibnamefont{Garrahan}}, \bibnamefont{and}
  \bibinfo{author}{\bibfnamefont{D.}~\bibnamefont{Chandler}},
  \bibinfo{journal}{Journal of Chemical Physics}
  \textbf{\bibinfo{volume}{123}}, \bibinfo{pages}{084509}
  (\bibinfo{year}{2005}), \urlprefix\url{http://dx.doi.org/10.1063/1.2001629}.

\bibitem[{\citenamefont{Kurchan et~al.}(1997)\citenamefont{Kurchan, Peliti, and
  Sellitto}}]{Kurchan1997}
\bibinfo{author}{\bibfnamefont{J.}~\bibnamefont{Kurchan}},
  \bibinfo{author}{\bibfnamefont{L.}~\bibnamefont{Peliti}}, \bibnamefont{and}
  \bibinfo{author}{\bibfnamefont{M.}~\bibnamefont{Sellitto}},
  \bibinfo{journal}{Europhysics Letters} \textbf{\bibinfo{volume}{39}},
  \bibinfo{pages}{365} (\bibinfo{year}{1997}),
  \urlprefix\url{http://dx.doi.org/10.1209/epl/i1997-00363-0}.

\bibitem[{\citenamefont{L{\'{e}}onard et~al.}(2007)\citenamefont{L{\'{e}}onard,
  Mayer, Sollich, Berthier, and Garrahan}}]{097Leonard2007}
\bibinfo{author}{\bibfnamefont{S.}~\bibnamefont{L{\'{e}}onard}},
  \bibinfo{author}{\bibfnamefont{P.}~\bibnamefont{Mayer}},
  \bibinfo{author}{\bibfnamefont{P.}~\bibnamefont{Sollich}},
  \bibinfo{author}{\bibfnamefont{L.}~\bibnamefont{Berthier}}, \bibnamefont{and}
  \bibinfo{author}{\bibfnamefont{J.}~\bibnamefont{Garrahan}},
  \bibinfo{journal}{Journal of Statistical Mechanics: Theory and Experiment} p.
  \bibinfo{pages}{P07017} (\bibinfo{year}{2007}),
  \urlprefix\url{http://dx.doi.org/10.1088/1742-5468/2007/07/P07017}.

\bibitem[{\citenamefont{Mayer and Sollich}(2007)}]{098Mayer2007}
\bibinfo{author}{\bibfnamefont{P.}~\bibnamefont{Mayer}} \bibnamefont{and}
  \bibinfo{author}{\bibfnamefont{P.}~\bibnamefont{Sollich}},
  \bibinfo{journal}{Journal of Physics A: Mathematical and Theoretical}
  \textbf{\bibinfo{volume}{40}}, \bibinfo{pages}{5823} (\bibinfo{year}{2007}),
  \urlprefix\url{http://dx.doi.org/10.1088/1751-8113/40/22/005}.

\bibitem[{\citenamefont{Toninelli and Biroli}(2007)}]{083Toninelli2007}
\bibinfo{author}{\bibfnamefont{C.}~\bibnamefont{Toninelli}} \bibnamefont{and}
  \bibinfo{author}{\bibfnamefont{G.}~\bibnamefont{Biroli}},
  \bibinfo{journal}{Journal of Statistical Physics}
  \textbf{\bibinfo{volume}{130}}, \bibinfo{pages}{83} (\bibinfo{year}{2007}),
  \urlprefix\url{http://dx.doi.org/10.1007/s10955-007-9420-z}.

\bibitem[{\citenamefont{Biroli and Toninelli}(2008)}]{084Biroli2008}
\bibinfo{author}{\bibfnamefont{G.}~\bibnamefont{Biroli}} \bibnamefont{and}
  \bibinfo{author}{\bibfnamefont{C.}~\bibnamefont{Toninelli}},
  \bibinfo{journal}{The European Physical Journal B}
  \textbf{\bibinfo{volume}{64}}, \bibinfo{pages}{567} (\bibinfo{year}{2008}),
  \urlprefix\url{http://dx.doi.org/10.1140/epjb/e2008-00029-9}.

\bibitem[{\citenamefont{Teomy and Shokef}(2012)}]{115Teomy2012}
\bibinfo{author}{\bibfnamefont{E.}~\bibnamefont{Teomy}} \bibnamefont{and}
  \bibinfo{author}{\bibfnamefont{Y.}~\bibnamefont{Shokef}},
  \bibinfo{journal}{Physical Review E} \textbf{\bibinfo{volume}{86}},
  \bibinfo{pages}{051133} (\bibinfo{year}{2012}),
  \urlprefix\url{http://dx.doi.org/10.1103/PhysRevE.86.051133}.

\bibitem[{\citenamefont{Teomy and Shokef}(2014{\natexlab{a}})}]{082Teomy2014}
\bibinfo{author}{\bibfnamefont{E.}~\bibnamefont{Teomy}} \bibnamefont{and}
  \bibinfo{author}{\bibfnamefont{Y.}~\bibnamefont{Shokef}},
  \bibinfo{journal}{Physical Review E} \textbf{\bibinfo{volume}{89}},
  \bibinfo{pages}{032204} (\bibinfo{year}{2014}{\natexlab{a}}),
  \urlprefix\url{http://dx.doi.org/10.1103/PhysRevE.89.032204}.

\bibitem[{\citenamefont{Ghosh et~al.}(2014)\citenamefont{Ghosh, Teomy, and
  Shokef}}]{085Ghosh2014}
\bibinfo{author}{\bibfnamefont{A.}~\bibnamefont{Ghosh}},
  \bibinfo{author}{\bibfnamefont{E.}~\bibnamefont{Teomy}}, \bibnamefont{and}
  \bibinfo{author}{\bibfnamefont{Y.}~\bibnamefont{Shokef}},
  \bibinfo{journal}{Europhysics Letters} \textbf{\bibinfo{volume}{106}},
  \bibinfo{pages}{16003} (\bibinfo{year}{2014}),
  \urlprefix\url{http://dx.doi.org/10.1209/0295-5075/106/16003}.

\bibitem[{\citenamefont{Corberi and Cugliandolo}(2009)}]{Corberi2009}
\bibinfo{author}{\bibfnamefont{F.}~\bibnamefont{Corberi}} \bibnamefont{and}
  \bibinfo{author}{\bibfnamefont{L.~F.} \bibnamefont{Cugliandolo}},
  \bibinfo{journal}{Journal of Statistical Mechanics: Theory and Experiment}
  \textbf{\bibinfo{volume}{2009}}, \bibinfo{pages}{P09015}
  (\bibinfo{year}{2009}),
  \urlprefix\url{http://dx.doi.org/10.1088/1742-5468/2009/09/P09015}.

\bibitem[{\citenamefont{Fielding}(2002)}]{Fielding2002}
\bibinfo{author}{\bibfnamefont{S.}~\bibnamefont{Fielding}},
  \bibinfo{journal}{Physical Review E} \textbf{\bibinfo{volume}{66}},
  \bibinfo{pages}{016103} (\bibinfo{year}{2002}),
  \urlprefix\url{http://dx.doi.org/10.1103/PhysRevE.66.016103}.

\bibitem[{\citenamefont{Sellitto}(2008)}]{Sellitto2008}
\bibinfo{author}{\bibfnamefont{M.}~\bibnamefont{Sellitto}},
  \bibinfo{journal}{Physical Review Letters} \textbf{\bibinfo{volume}{101}},
  \bibinfo{pages}{048301} (\bibinfo{year}{2008}),
  \urlprefix\url{http://dx.doi.org/10.1103/PhysRevLett.101.048301}.

\bibitem[{\citenamefont{Shokef and Liu}(2010)}]{041Shokef2010}
\bibinfo{author}{\bibfnamefont{Y.}~\bibnamefont{Shokef}} \bibnamefont{and}
  \bibinfo{author}{\bibfnamefont{A.}~\bibnamefont{Liu}},
  \bibinfo{journal}{Europhysics Letters} \textbf{\bibinfo{volume}{90}},
  \bibinfo{pages}{26005} (\bibinfo{year}{2010}),
  \urlprefix\url{http://dx.doi.org/10.1209/0295-5075/90/26005}.

\bibitem[{\citenamefont{Turci et~al.}(2012)\citenamefont{Turci, Pitard, and
  Sellitto}}]{130Turci2012a}
\bibinfo{author}{\bibfnamefont{F.}~\bibnamefont{Turci}},
  \bibinfo{author}{\bibfnamefont{E.}~\bibnamefont{Pitard}}, \bibnamefont{and}
  \bibinfo{author}{\bibfnamefont{M.}~\bibnamefont{Sellitto}},
  \bibinfo{journal}{Physical Review E} \textbf{\bibinfo{volume}{86}},
  \bibinfo{pages}{031112} (\bibinfo{year}{2012}),
  \urlprefix\url{http://dx.doi.org/10.1103/PhysRevE.86.031112}.

\bibitem[{\citenamefont{Sellitto}(2002)}]{126Sellitto2002}
\bibinfo{author}{\bibfnamefont{M.}~\bibnamefont{Sellitto}},
  \bibinfo{journal}{Physical Review E} \textbf{\bibinfo{volume}{65}},
  \bibinfo{pages}{020101} (\bibinfo{year}{2002}),
  \urlprefix\url{http://dx.doi.org/10.1103/PhysRevE.65.020101}.

\bibitem[{\citenamefont{Gon{\c{c}}alves
  et~al.}(2009)\citenamefont{Gon{\c{c}}alves, Landim, and
  Toninelli}}]{127Goncalves2009}
\bibinfo{author}{\bibfnamefont{P.}~\bibnamefont{Gon{\c{c}}alves}},
  \bibinfo{author}{\bibfnamefont{C.}~\bibnamefont{Landim}}, \bibnamefont{and}
  \bibinfo{author}{\bibfnamefont{C.}~\bibnamefont{Toninelli}},
  \bibinfo{journal}{Annales de l'institut Henri Poincare (B) Probability and
  Statistics} \textbf{\bibinfo{volume}{45}}, \bibinfo{pages}{887}
  (\bibinfo{year}{2009}), \urlprefix\url{http://dx.doi.org/10.1214/09-AIHP210}.

\bibitem[{\citenamefont{Karabanov et~al.}(2015)\citenamefont{Karabanov,
  Wi{\'{s}}niewski, Lesanovsky, and K{\"{o}}ckenberger}}]{Karabanov2015}
\bibinfo{author}{\bibfnamefont{a.}~\bibnamefont{Karabanov}},
  \bibinfo{author}{\bibfnamefont{D.}~\bibnamefont{Wi{\'{s}}niewski}},
  \bibinfo{author}{\bibfnamefont{I.}~\bibnamefont{Lesanovsky}},
  \bibnamefont{and}
  \bibinfo{author}{\bibfnamefont{W.}~\bibnamefont{K{\"{o}}ckenberger}},
  \bibinfo{journal}{Physical Review Letters} \textbf{\bibinfo{volume}{115}},
  \bibinfo{pages}{020404} (\bibinfo{year}{2015}),
  \urlprefix\url{http://link.aps.org/doi/10.1103/PhysRevLett.115.020404}.

\bibitem[{\citenamefont{Spitzer}(1970)}]{068Spitzer1970}
\bibinfo{author}{\bibfnamefont{F.}~\bibnamefont{Spitzer}},
  \bibinfo{journal}{Advances in Mathematics} \textbf{\bibinfo{volume}{5}},
  \bibinfo{pages}{246} (\bibinfo{year}{1970}),
  \urlprefix\url{http://dx.doi.org/10.1016/0001-8708(70)90034-4}.

\bibitem[{\citenamefont{Toninelli et~al.}(2004)\citenamefont{Toninelli, Biroli,
  and Fisher}}]{071Toninelli2004a}
\bibinfo{author}{\bibfnamefont{C.}~\bibnamefont{Toninelli}},
  \bibinfo{author}{\bibfnamefont{G.}~\bibnamefont{Biroli}}, \bibnamefont{and}
  \bibinfo{author}{\bibfnamefont{D.}~\bibnamefont{Fisher}},
  \bibinfo{journal}{Physical Review Letters} \textbf{\bibinfo{volume}{92}},
  \bibinfo{pages}{185504} (\bibinfo{year}{2004}),
  \urlprefix\url{http://dx.doi.org/10.1103/PhysRevLett.92.185504}.

\bibitem[{\citenamefont{Arita et~al.}(2014)\citenamefont{Arita, Krapivsky, and
  Mallick}}]{125Arita2014}
\bibinfo{author}{\bibfnamefont{C.}~\bibnamefont{Arita}},
  \bibinfo{author}{\bibfnamefont{P.}~\bibnamefont{Krapivsky}},
  \bibnamefont{and} \bibinfo{author}{\bibfnamefont{K.}~\bibnamefont{Mallick}},
  \bibinfo{journal}{Physical Review E} \textbf{\bibinfo{volume}{90}},
  \bibinfo{pages}{052108} (\bibinfo{year}{2014}),
  \urlprefix\url{http:///dx.doi.org/10.1103/PhysRevE.90.052108}.

\bibitem[{\citenamefont{Nakazato and Kitahara}(1980)}]{124Nakazato1980}
\bibinfo{author}{\bibfnamefont{K.}~\bibnamefont{Nakazato}} \bibnamefont{and}
  \bibinfo{author}{\bibfnamefont{K.}~\bibnamefont{Kitahara}},
  \bibinfo{journal}{Progress of Theoretical Physics}
  \textbf{\bibinfo{volume}{64}}, \bibinfo{pages}{2261} (\bibinfo{year}{1980}),
  \urlprefix\url{http://dx.doi.org/10.1143/PTP.64.2261}.

\bibitem[{\citenamefont{Toninelli}(2003)}]{116Toninelli2003}
\bibinfo{author}{\bibfnamefont{C.}~\bibnamefont{Toninelli}}, Ph.D. thesis
  (\bibinfo{year}{2003}),
  \urlprefix\url{http://www.proba.jussieu.fr/~toninelli/tesidottorato.ps}.

\bibitem[{\citenamefont{Holroyd}(2003)}]{073Holroyd2003}
\bibinfo{author}{\bibfnamefont{A.}~\bibnamefont{Holroyd}},
  \bibinfo{journal}{Probability Theory and Related Fields}
  \textbf{\bibinfo{volume}{125}}, \bibinfo{pages}{195} (\bibinfo{year}{2003}),
  \urlprefix\url{http://dx.doi.org/10.1007/s00440-002-0239-x}.

\bibitem[{\citenamefont{Teomy and Shokef}(2014{\natexlab{b}})}]{072Teomy2014a}
\bibinfo{author}{\bibfnamefont{E.}~\bibnamefont{Teomy}} \bibnamefont{and}
  \bibinfo{author}{\bibfnamefont{Y.}~\bibnamefont{Shokef}},
  \bibinfo{journal}{Journal of Chemical Physics}
  \textbf{\bibinfo{volume}{141}}, \bibinfo{pages}{064110}
  (\bibinfo{year}{2014}{\natexlab{b}}),
  \urlprefix\url{http://dx.doi.org/10.1063/1.4892416}.

\bibitem[{\citenamefont{Weisstein}()}]{128WeissteinIncompleteBeta}
\bibinfo{author}{\bibfnamefont{E.~W.} \bibnamefont{Weisstein}},
  \emph{\bibinfo{title}{{Incomplete Beta Function. From MathWorld---A Wolfram
  Web Resource}}},
  \urlprefix\url{http://mathworld.wolfram.com/IncompleteBetaFunction.html}.

\bibitem[{\citenamefont{Aminov et~al.}(2015)\citenamefont{Aminov, Y., and
  M.}}]{129Aminov2015}
\bibinfo{author}{\bibfnamefont{A.}~\bibnamefont{Aminov}},
  \bibinfo{author}{\bibfnamefont{K.}~\bibnamefont{Y.}}, \bibnamefont{and}
  \bibinfo{author}{\bibfnamefont{K.}~\bibnamefont{M.}},
  \bibinfo{journal}{Physical Review Letters} \textbf{\bibinfo{volume}{114}},
  \bibinfo{pages}{230602} (\bibinfo{year}{2015}),
  \urlprefix\url{http://dx.doi.org/10.1103/PhysRevLett.114.230602}.

\bibitem[{\citenamefont{Derrida et~al.}(1993)\citenamefont{Derrida, Evans,
  Hakim, and Pasquier}}]{Derrida1993}
\bibinfo{author}{\bibfnamefont{B.}~\bibnamefont{Derrida}},
  \bibinfo{author}{\bibfnamefont{M.}~\bibnamefont{Evans}},
  \bibinfo{author}{\bibfnamefont{V.}~\bibnamefont{Hakim}}, \bibnamefont{and}
  \bibinfo{author}{\bibfnamefont{V.}~\bibnamefont{Pasquier}},
  \bibinfo{journal}{Journal of Physics A: Mathematical and General}
  \textbf{\bibinfo{volume}{26}}, \bibinfo{pages}{1493} (\bibinfo{year}{1993}),
  \urlprefix\url{http://dx.doi.org/10.1088/0305-4470/26/7/011}.

\bibitem[{\citenamefont{Krapivsky and Meerson}(2012)}]{Krapivsky2012}
\bibinfo{author}{\bibfnamefont{P.}~\bibnamefont{Krapivsky}} \bibnamefont{and}
  \bibinfo{author}{\bibfnamefont{B.}~\bibnamefont{Meerson}},
  \bibinfo{journal}{Physical Review E} \textbf{\bibinfo{volume}{86}},
  \bibinfo{pages}{031106} (\bibinfo{year}{2012}),
  \urlprefix\url{http://dx.doi.org/10.1103/PhysRevE.86.031106}.

\bibitem[{\citenamefont{Segall et~al.}(2016)\citenamefont{Segall, Teomy, and
  Shokef}}]{086Segall2016}
\bibinfo{author}{\bibfnamefont{N.}~\bibnamefont{Segall}},
  \bibinfo{author}{\bibfnamefont{E.}~\bibnamefont{Teomy}}, \bibnamefont{and}
  \bibinfo{author}{\bibfnamefont{Y.}~\bibnamefont{Shokef}},
  \bibinfo{journal}{Journal of Statistical Mechacnics: Theory and Experiment}
  \textbf{\bibinfo{volume}{2016}}, \bibinfo{pages}{054051}
  (\bibinfo{year}{2016}),
  \urlprefix\url{http://dx.doi.org/10.1088/1742-5468/2016/05/054051}.

\end{thebibliography}

\end{document}